\documentclass[sigconf]{acmart}
\pdfoutput=1

\usepackage[latin1]{inputenc}
\usepackage[T1]{fontenc}
\usepackage{xspace}
\usepackage{graphicx}
\usepackage{amssymb}
\usepackage{enumitem}
\setitemize{noitemsep,topsep=0pt,parsep=0pt,partopsep=0pt}
\setlist[itemize,1]{leftmargin=\dimexpr 26pt-10pt}

\usepackage{balance} 
\usepackage{float}

\newcommand{\skpline}{\vspace{-0.2cm}}

\newcommand {\eg}{e.g.,\xspace}
\newcommand {\ie}{i.e.,\xspace}

\newcommand{\wrt}{w.r.t.\xspace}

\AtBeginDocument{%
  \providecommand\BibTeX{{%
    \normalfont B\kern-0.5em{\scshape i\kern-0.25em b}\kern-0.8em\TeX}}}

\renewcommand\footnotetextcopyrightpermission[1]{}

\begin{document}
\fancyhead{}

\title{Privacy Protection in Distributed Fingerprint-based Authentication}


\author{Swe Geng}
\authornote{Joint work with ABB Corporate Research.}
\affiliation{%
  \institution{ETH Z\"urich}
  \city{Z\"urich}
  \country{Switzerland}
}

\email{swe.geng@alumni.ethz.ch}

\author{Georgia Giannopoulou}
\affiliation{%
  \institution{ABB Corporate Research}
  \city{Baden-D\"attwil}
  \country{Switzerland}}
\email{georgia.giannopoulou@ch.abb.com}

\author{Ma\"elle Kabir-Querrec}
\affiliation{%
  \institution{ABB Corporate Research}
  \city{Baden-D\"attwil}
  \country{Switzerland}}
\email{maelle.kabir-querrec@ch.abb.com}


\pagestyle{plain} 

\begin{abstract}
 Biometric authentication is getting increasingly popular due to the convenience of using unique individual traits, 
such as fingerprints, palm veins, irises. 
Especially fingerprints are widely used nowadays due to the availability and 
low cost of fingerprint scanners. To avoid identity theft or impersonation, fingerprint data is typically
stored locally, \eg in a trusted hardware module, in a single device that is used for user enrollment and authentication. 
Local storage, however, limits the ability to implement distributed applications,
in which users can enroll their fingerprint once and use it to access multiple physical locations and mobile applications afterwards.

In this paper, we present a distributed authentication system that stores fingerprint data in a server or cloud infrastructure in a privacy-preserving way. 
Multiple devices can be connected and perform user enrollment or verification.
To secure the privacy and integrity of sensitive data, we employ a cryptographic construct called fuzzy vault. 
We highlight challenges in implementing fuzzy vault-based authentication, for which we propose and compare alternative solutions.
We conduct a security analysis of our biometric cryptosystem, and as a proof of concept,
we build an authentication system for access control using resource-constrained devices (Raspberry Pis) 
connected to fingerprint scanners and the Microsoft Azure cloud environment. 
Furthermore, we evaluate the fingerprint matching algorithm against the 
well-known FVC2006 database and show that it can achieve comparable accuracy to widely-used
matching techniques that are not designed for privacy, while remaining efficient with an authentication time of few seconds.

\end{abstract}


\maketitle

\section{Introduction}
\textbf{Motivation.} 
Digital identities are essential to user applications in our interconnected world. Digital identity theft or impersonation attacks 
can compromise the affected individuals' privacy and lead to severe consequences, such as financial loss or legal violations.
This risk creates the need for reliable authentication techniques. 
Biometric authentication offers the benefit of high reliability,
since, unlike smart cards or text passwords, biometrics cannot be lost or forgotten, 
and biometric forgery or theft is very difficult. 

As a result, fingerprint authentication is frequently used \eg in smart phones, which store and manage fingerprint data locally in trusted execution environments (TEE),
such as Trusty TEE~\cite{trusty} and Secure Enclave~\cite{enclave} for Android and Apple devices, respectively.
The practice of secure local storage protects sensitive data from leaking to external attackers, however it also prevents the development of (efficient) distributed
biometric authentication. A use case of distributed authentication which we study in this work is a fingerprint-based access control system for
multiple buildings of an international company that are equipped with fingerprint scanners at their entrances.
The scanners allow employees to authenticate themselves for access to the respective buildings. A distributed authentication system eliminates the need to
enroll fingerprints at each site separately, by using a one-time sign up instead and enabling access to multiple locations afterwards. For this, the employees only need to swipe 
their fingers without worrying about forgetting passwords, losing a smart card or relying on extra equipment like a smart phone.

Despite its usability and convenience, distributed authentication is still not widely accepted due to the high
risk of biometrics being stolen when stored online in untrusted environments~\cite{prabhakar2003biometric}. 
In case of fingerprint data leak, not only can an identity 
be misused across several locations/applications, but the same finger can never be reused for authentication. 
Revocation of fingerprint data is very challenging as each individual typically possesses 10 different fingerprints.
Our work aims at developing a distributed fingerprint-based authentication system that protects the users' sensitive data.

To protect biometric data, most existing works integrate biometric recognition with cryptographic techniques, 
following one of two approaches: \textit{cancelable biometrics} or \textit{biometric cryptosystems}~\cite{rathgeb2011survey}. 
Cancelable biometrics include techniques that convert the biometric into a transformed domain, where the comparison for user verification also takes place. 
The inversion of the transformed biometrics must be infeasible, \ie an attacker cannot retrieve the original biometric 
from the transformed domain. In contrast, a biometric cryptosystem binds a digital key to a biometric or generates a digital key directly from given biometrics, 
yielding key-binding or key-generation schemes, respectively. 

An established key-binding biometric cryptosystem is \textit{fuzzy vault}~\cite{juels2006fuzzy}. 
Fuzzy vault is a secure data structure which binds a secret key with a biometric (in our case, fingerprint).
The key can only be extracted by a fingerprint capture sufficiently similar to the one bound to it so that
only the owner can unlock the fuzzy vault and retrieve the fingerprint data.
Fuzzy vault provides strong security guarantees by protecting the bound fingerprint with a lot of random noise. 
Extraction of the fingerprint is immensely difficult without already having a 
capture of the same finger, which reduces the risk of identity theft significantly compared to cancelable 
biometrics approaches~\cite{juels2006fuzzy}.
Furthermore, prototype implementations of fuzzy vault show promising results in terms of accuracy in fingerprint recognition~\cite{nandakumar2007fingerprint}. 
Due to its strong security and accuracy, 
we base our authentication system on fuzzy vault and present its first deployment in a real-world distributed setting. \\ \skpline

\noindent
\textbf{Contributions.}
This paper addresses the problem of developing a secure and usable distributed biometric authentication system.
The quality of a biometric system is typically evaluated \wrt its false match rate (FMR) and false non-match rate (FNMR), 
also referred to as matching accuracy~\cite{maltoni2009handbook}.
For security, we aim to minimize the FMR of the solution.
For usability, we aim to minimize the FNMR as well as the time needed for a successful authentication.
%
Our main contributions towards these goals are listed below:
\begin{itemize}
	\item We propose a distributed biometric cryptosystem in which user fingerprint data is stored in the cloud 
	and several connected devices can enroll and authenticate users without maintaining local fingerprint copies.
	Our solution is based on a well-known technique with proven security properties, \ie fuzzy vault~\cite{juels2006fuzzy}.  
	\item We highlight implementation challenges of certain primitives,
	propose alternative solutions and analyze their trade-offs \wrt security and usability.
	Based on our analysis, we recommend using (i) polynomial operations in a Galois field to allow exact interpolation during the fuzzy vault decoding phase, 
   (ii) geometric hashing for the alignment of fingerprint templates which improves security and accuracy compared to other 
   alignment methods without relying on public helper data,
   (iii) a custom configuration of fuzzy vault matching thresholds depending on the requirements of each individual use case.
	\item We analyze the security of the distributed biometric cryptosystem and estimate breach probabilities.
	\item We develop a proof-of-concept implementation with commercial fingerprint scanners connected with embedded devices (Raspberry Pis) and a Microsoft
	Azure cloud.
	\item We evaluate our system extensively against the well-known FVC2006 fingerprint database and demonstrate its applicability to real-world scenarios.
	The system provides comparable accuracy to widely-used methods for fingerprint recognition that were not designed for privacy,
	with acceptable authentication times of less than 3 seconds on regular computers. It also improves upon existing biometric cryptosystems
	in terms of security and/or accuracy.	\\  \skpline
\end{itemize}
Note that this paper is an extended version of~\cite{geng2019privacy}.

\noindent
\textbf{Outline.}
In the remainder of the paper, \autoref{sec:related-work} presents related work. 
\autoref{sec:system-threat-model} presents our system and threat model assumptions. 
\autoref{sec:implementation} contains an overview of the fuzzy vault structure, implementation challenges that we encountered and the chosen solutions with their 
respective trade-offs.
\autoref{sec:distributed-fingerprint-application} details the implementation of a distributed fingerprint-based authentication application for access control
in a multinational company. 
\autoref{sec:attack-genuine-vault-pairs} presents the security analysis and \autoref{sec:results} the empirical evaluation
of our cryptosystem. \autoref{sec:conclusion} concludes the paper.

\section{Related Work}\label{sec:related-work}
\textbf{Fingerprint recognition.}
Due to the increasing use of biometric authentication, fingerprint recognition 
is a well-studied research topic. 
Fingerprint recognition algorithms extract a mathematical representation from a fingerprint image, called template, and use it to verify a match between two fingerprints. 
Matching requires the comparison of several features of the print pattern,
such as the valleys and ridges of the finger skin.
Most recognition algorithms focus on features from minutiae,
which are points defined as ridge endings (where a valley splits) or ridge bifurcations (where a
ridge splits), while other techniques include correlation-based or ridge feature-based matching~\cite{maltoni2009handbook}.
Commonly used minutiae-based matching techniques include the NIST Biometric Image Software (NBIS)~\cite{ko2007user}, Verifinger~\cite{website:verifinger} and 
Minutia Cylinder-Code (MCC)~\cite{cappelli2010minutia}. These algorithms only match fingerprint templates without guaranteeing the security of the templates. \\ \skpline

\noindent
\textbf{Fingerprint template protection.}
In literature, the two fundamental approaches in template protection are cancelable biometrics and biometric cryptosystems~\cite{rathgeb2011survey,maltoni2009handbook}. 
In cancelable biometrics, feature transformation techniques, including non-invertible transforms and salting, transform an unprotected enrollment template 
into a protected one with a respective transformation function. During verification the probe or verification template is also transformed in the same way and comparison
is performed in the transformed space. Ferrara et al.~\cite{ferrara2012noninvertible} implemented the Protected Minutia Cylinder-Code (P-MCC) based on 
MCC~\cite{cappelli2010minutia} using non-invertible transforms. 
In contrast, biometric cryptosystems either generate a cryptographic key from a fingerprint (key-generation) or secure a key using a fingerprint 
(key-binding). A prominent example of key-generation techniques is the fuzzy extractor~\cite{dodis2004fuzzy}, where a cryptographic key is generated from noisy fingerprint data. 
As for key-binding biometric cryptosystems, fuzzy vault~\cite{juels2006fuzzy} is among the most known techniques. We detail its implementation in 
\autoref{subsec:fuzzy-vault-detailed-concept}.

We chose to build our biometric cryptosystem upon fuzzy vault because of its proven security guarantees~\cite{juels2006fuzzy} and the high matching accuracy compared to 
other template protection approaches \cite{nandakumar2007fingerprint}. Additionally,
since our biometric cryptosystem targets distributed applications, the fuzzy vault is advantageous as the fingerprint image or template 
can be directly deleted after locally creating a fuzzy vault, thus reducing the attacker surface (the raw template is not stored in any local device). 
Despite the extensive literature on fuzzy vaults, to the best of our knowledge, this cryptosystem has not been employed in industrial applications
or evaluated in real-world settings. Existing fingerprint-based applications that have been based on fuzzy vault, like the ones by Uludag et al.~\cite{uludag2005fuzzy},
Nandakumar et al.~\cite{nandakumar2007fingerprint} or Theodorakis~\cite{theodorakis2018secure}, are implemented in MATLAB and evaluated with public
fingerprint databases. Li et al.~\cite{li2010alignment} have implemented an alignment-free fuzzy vault approach in C++. 
Our implementation is the first to use fuzzy vault in a distributed setting and also to evaluate it in a real-world setting
with actual fingerprint devices.
The major differences \wrt previous implementations lies in our approach for fingerprint alignment, as detailed in \autoref{subsec:fingerprint-authentication-algorithm}.\\ \skpline

\noindent
\textbf{Fingerprint alignment.}
Alignment of fingerprint templates prior to matching is a great challenge for most fingerprint identification systems, incl.~fuzzy vault based systems. 
Jeffers et al.~\cite{jeffers2007fingerprint} tested three known alignment approaches with five nearest neighbor, Voronoi neighbors and triangle based structures for fuzzy vault.
Uludag et al.~\cite{uludag2006securing} used an orientation-field based approach that relies on public helper data. 
Li et al.~\cite{li2016security} introduced pair-polar structures that facilitate alignment but need changing functions to prevent information leakage. 
Chung et al.~\cite{chung2005automatic} applied geometric hashing in the fuzzy vault setting, which is computationally more expensive than other methods 
but uses global features of the fingerprint template and does not need public helper data. 
In our biometric cryptosystem we use the geometric hashing approach of Chung et al.~\cite{chung2005automatic} due to its favorable performance compared to other alignment methods. Moreover, this approach does not require any helper data which strengthens the system security.
\section{System and Threat Model}
\label{sec:system-threat-model}

\subsection{System Model}
\label{subsec:system-model}
\autoref{fig:system-threat-model} depicts our distributed fingerprint authentication system model. 
This features an arbitrary number of local computers equipped with fingerprint sensors (FPS). 
The computers are connected to the Internet over one or multiple remote servers in the cloud.
The system supports the following operations:
\begin{enumerate}
	\item Capture fingerprint raw image 
	\item Create template and secure data structure from raw image 
	\item Store secure data structure of fingerprint 
	\item Perform fingerprint matching for a probe fingerprint 
\end{enumerate}

The three first operations are required for the \textit{enrollment} of a fingerprint to the system, 
which happens once.
After an image of the fingerpint is captured by a FPS (1), the raw image is processed, \eg binarized,
at the connected computer to enable feature extraction (2).
A template is a mathematical representation of the fingerprint features (in our system, minutiae) 
and thus, contains sensitive data. To protect the template, we convert it to a secure data structure 
(in our system, fuzzy vault) and store it in a database (3).
The forth operation uses the stored data structure for user \textit{verification}. This entails extracting the
template from a newly captured (probe) fingerprint and comparing it to the enrolled template of the claimed user.
The matching algorithm calculates a similarity score and decides whether it is a match or non-match
based on a similarity threshold.
This operation can be executed multiple times.

Our system allows enrolling and verifying a fingerprint at any local computer. For instance,
a user can enroll a fingerprint at the computer labeled $2)$ in \autoref{fig:system-threat-model} and then authenticate at another location, \eg the local computers labeled $4)$. The fingerprint database (in our system, collection of fuzzy vaults) is stored in the cloud.

\subsection{Threat Model}
\label{subsec:threat-model}
\begin{figure}[t]
	\centering
	\includegraphics[width=\linewidth]{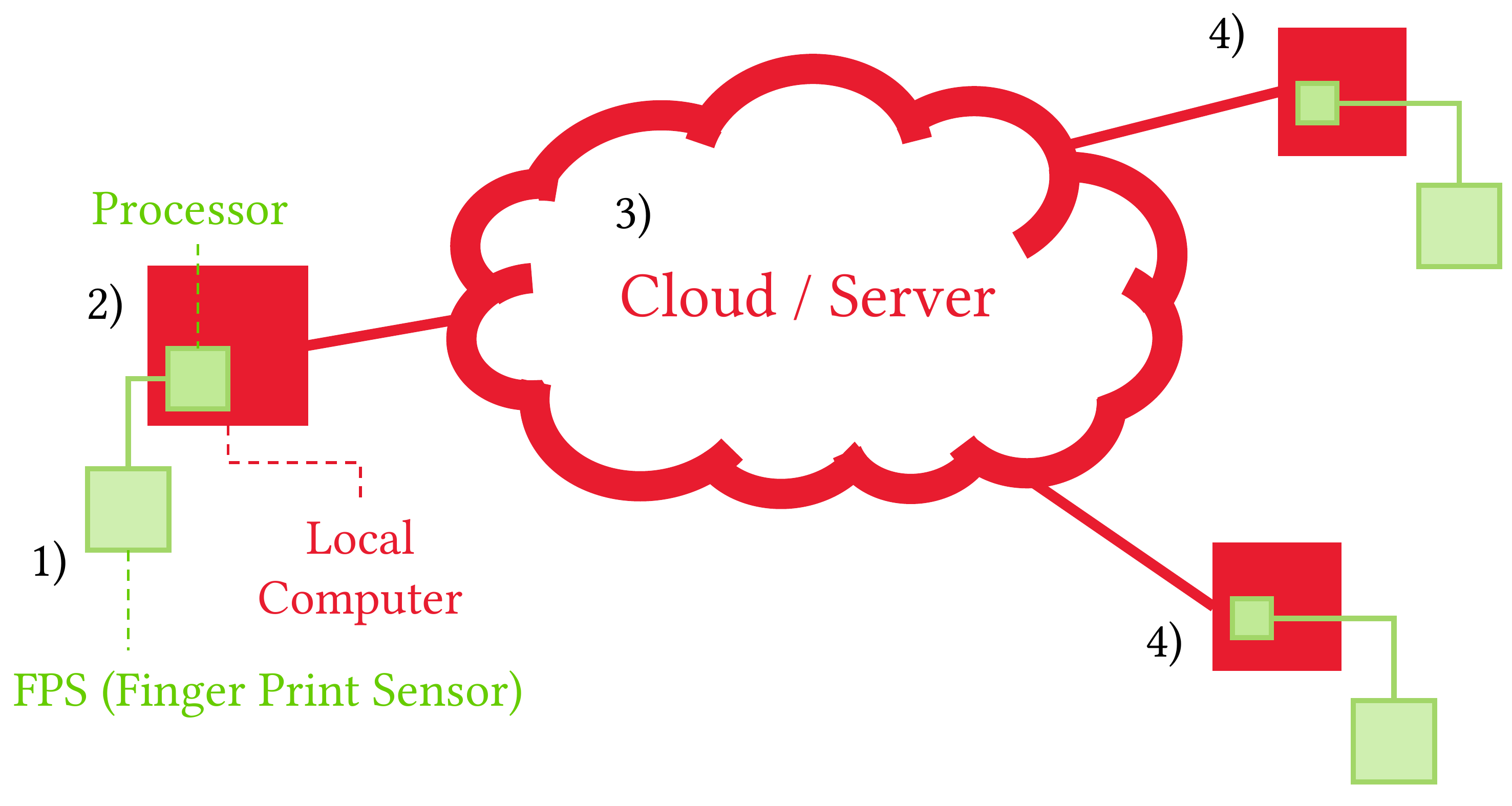}
	\caption{System and threat model. The components in green (red) are assumed to be trusted (non-trusted).}
	\label{fig:system-threat-model}
\end{figure}

In our system, the attacker either has the intention of extracting valuable assets or wants to gain unauthorized access by fooling the biometric system. 
We consider that the attacker is successful if she can determine the genuine minutiae in a fuzzy vault or if she can successfully authenticate against a fuzzy vault without the appropriate biometrics. In the first case, the attacker can perform identity theft with the extracted biometrics. In the second case, the attacker can gain unauthorized access and perform identity theft with the specific fuzzy vault as well.

\autoref{fig:system-threat-model} depicts the trusted parts of the system in green and the non-trusted ones in red, respectively.
Note that we do not consider attacks to the physical part of the biometric system. For instance, spoofing the fingerprint sensor with a mold is out of the scope of this work. 
We also do not investigate side channel attacks on the physical device. We assume that the FPS is directly connected to a trusted processor and our application is running in a secure execution environment. Finally,
the injection of fake profiles to the system to achieve a successful authentication is out of scope, as our work focuses on the protection and privacy of genuine users' biometrics.

Neither the components outside the processor on the local computer are trusted, nor the connections to the server and the server itself. The attacker can eavesdrop, alter messages or feed in messages to perform attacks on the channels. We assume that the server is honest but curious, \ie when asked for a specific fingerprint data structure, the server will send the correct data. 
Denial of service attacks are not considered. In case of such attacks, the worst-case scenario is that an authorized user will not be authenticated, but biometric 
data privacy will not be compromised. 

We assume that the attacker cannot tamper with the internals of our algorithm, which is only running in the secure execution environment of the local computer. However, the attacker can have knowledge over the used parameters in the algorithm except of the chosen secret. Moreover, we presume that the attacker can present fingerprint templates of his/her own to our algorithm without having to spoof the FPS directly. This allows a brute-force attack with fingerprint templates to attempt unlocking a fuzzy vault. We consider this scenario more realistic than forcing the attacker to only use the FPS to interact with the whole system.
\thispagestyle{empty}
\section{Biometric Cryptosystem}
\label{sec:implementation}

To implement our distributed biometric cryptosystem, we start by developing a fingerprint authentication algorithm
which uses a secure construct, \ie fuzzy vault, for protecting sensitive data. 
\autoref{subsec:fuzzy-vault-detailed-concept} presents an overview of fuzzy vault~\cite{juels2006fuzzy}. 
\autoref{subsec:fingerprint-authentication-algorithm} details our concrete implementation of the fingerprint authentication algorithm
and \autoref{subsec:implementation-challenges-contributions} proposes alternative solutions to implement certain crypto-primitives with their trade-offs.

\subsection{Fuzzy Vault Concept}
\label{subsec:fuzzy-vault-detailed-concept}

The principle of fuzzy vault was first introduced by Juels et al.~\cite{juels2006fuzzy} and is a cryptographic construction which ``locks'' a secret key $K$ using a set of elements $A$. The secret key $K$ can then only be ``unlocked'' by a set of elements $B$ which is sufficiently similar to set $A$. Since the unlocking criterion is the set similarity, the order of elements is irrelevant. To obscure the genuine points (from set $A$) from a possible attacker, a number of chaff points are added to the fuzzy vault. Those chaff points are randomly generated and cannot be distinguished from genuine points by an attacker. When trying to retrieve the secret, the goal is to determine as many points in set $A$ as possible with the help of set $B$ (given their similarity).

This approach is suitable for a biometric application like fingerprint authentication, as the various captures of a fingerprint always differ because of distortions or misalignment. In this case, sets $A$ and $B$ can be constructed from two fingerprint captures using the \emph{probe} and \emph{gallery} minutiae, respectively. The probe fingerprint is used to generate a fuzzy vault (at every authentication request) and the gallery fingerprint (stored at enrollment) is used to verify and unlock the fuzzy vault and authenticate the user. The goal is to use the probe minutiae to detect as many genuine gallery minutiae in the vault as possible.

\begin{figure}[t]
	\centering
	\includegraphics[width=\linewidth]{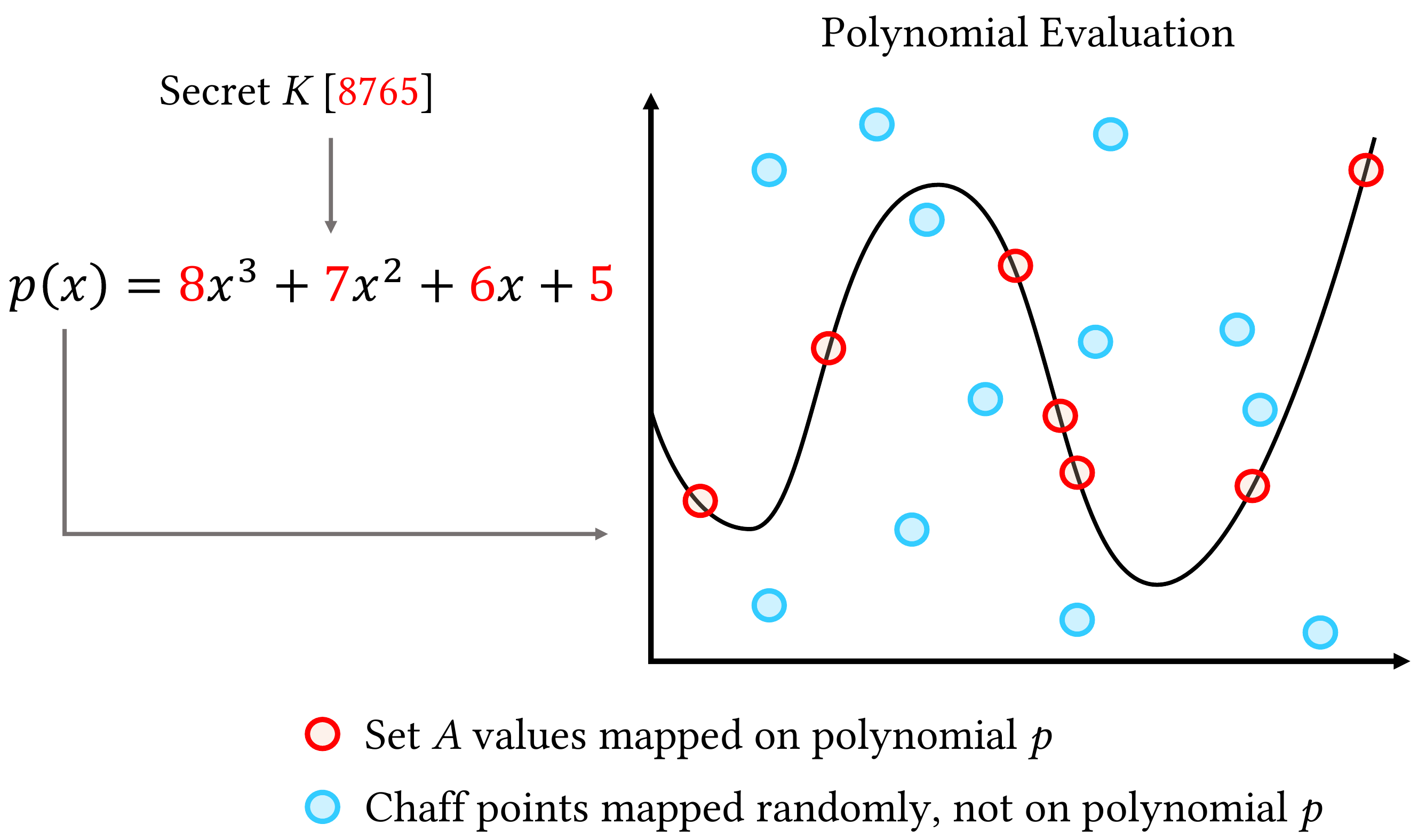}
	\caption{Fuzzy vault: Encoding phase.}
	\label{fig:encode-polynomial}
\end{figure}
In the encoding phase, using set $A$, the secret key $K$ is embedded in a polynomial $p$, typically as its coefficients. An example is given in \autoref{fig:encode-polynomial} with a simple secret $K$. The elements in set $A$ are then treated as distinct coordinate values and mapped onto $p$ as can be seen in red. The chaff points are chosen such that their mappings do not lie on the polynomial $p$.

\begin{figure}[t]
	\centering
	\includegraphics[width=\linewidth]{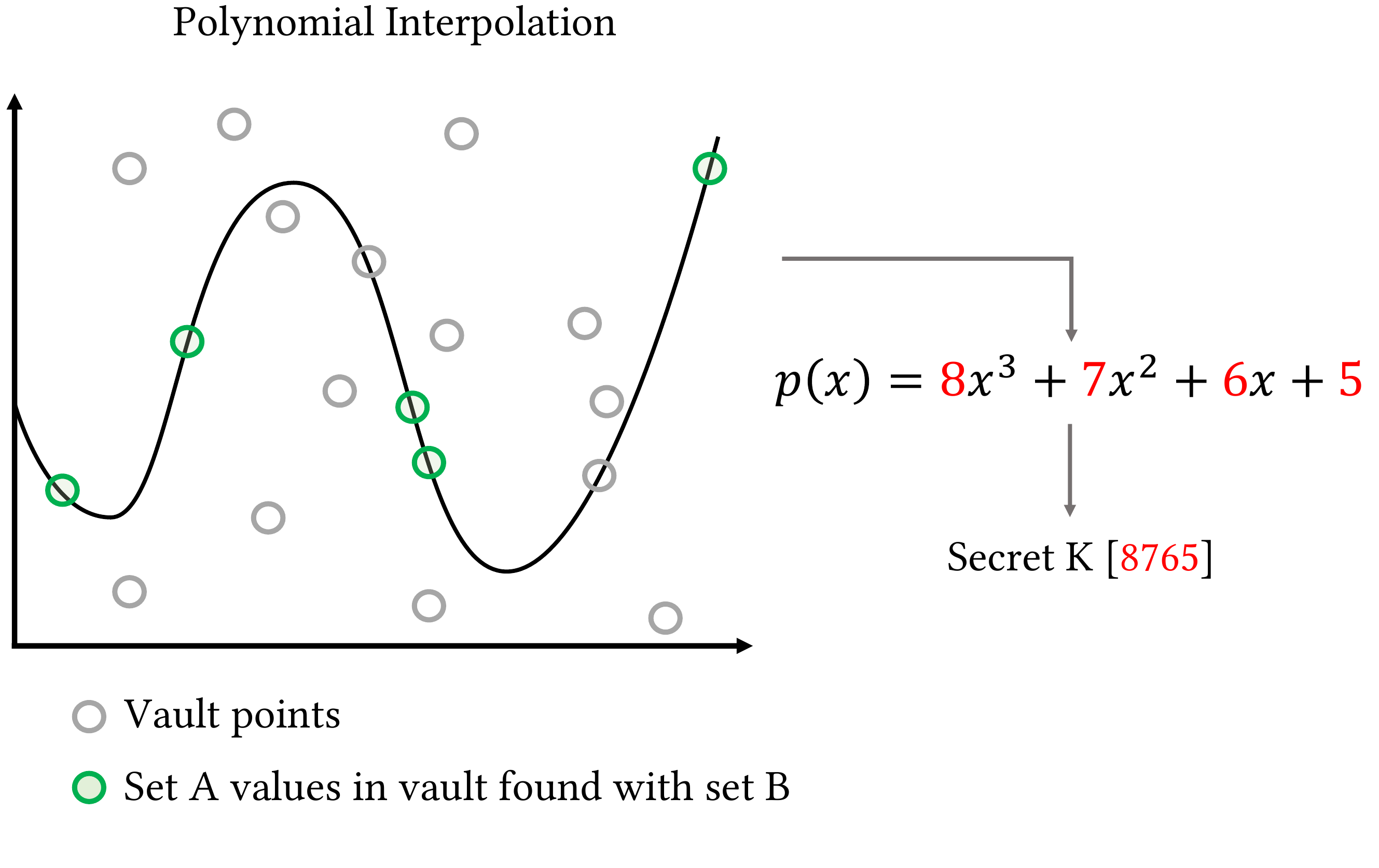}
	\caption{Fuzzy vault: Decoding phase.}
	\label{fig:decode-polynomial}
\end{figure}

In the decoding phase, using set $B$, several points are found that lie on polynomial $p$ (from set $A$). These are depicted in green in \autoref{fig:decode-polynomial}. The secret $K$ can be derived with polynomial interpolation if $n + 1$ elements in $A$ are found, where $n$ is the polynomial degree of $p$. If the correct secret is retrieved, the user is successfully authenticated since the probe fingerprint unlocks the fuzzy vault.

The fuzzy vault concept has been adopted for minutiae-based fingerprint authentication~\cite{uludag2005fuzzy,nandakumar2007fingerprint}.
The original work~\cite{juels2006fuzzy} suggested error-correction codes for bridging the differences between sets $A$ and $B$. 
Uludag et al.~\cite{uludag2005fuzzy} identified the difficulty in applying error correction to biometrics 
because of large variations that exceed the error-correction abilities of such codes. 
They proposed an algorithm based on cyclic redundancy check and Lagrange interpolation that decodes many candidate secrets, 
resulting in more interpolation attempts. Nandakumar et al.~\cite{nandakumar2007fingerprint} further improved the algorithm by 
considering non-linear distortion in fingerprints and also, orientation in addition to the location of the minutiae.
The algorithm of~\cite{uludag2005fuzzy} has been widely used in literature. We build and improve upon it, especially \wrt
fingerprint alignment and security, as described in the following section.

\subsection{Fingerprint Authentication}
\label{subsec:fingerprint-authentication-algorithm}
To use the fuzzy vault concept for fingerprint authentication, vault encoding and decoding need to be implemented. 
In the vault encoding phase, the vault is constructed with the minutiae from a fingerprint gallery template of a given user and a secret key. 
The fuzzy vault can be used afterwards to authenticate the given user. 
Vault decoding uses the previously generated fuzzy vault and the minutiae from a probe template of the user to attempt retrieval of the secret key. 
After key retrieval, an integrity check like a cyclic redundancy check (CRC) is conducted to determine if the correct secret key has been found. 
In this case, the secret key does not need to be stored anywhere to check if there is a match. 

Our implementation of vault encoding and decoding follows similar principles as previous works~\cite{uludag2005fuzzy,nandakumar2007fingerprint} 
and is illustrated in \autoref{fig:vault-encoding} and~\ref{fig:vault-decoding}, respectively. 
Compared to~\cite{uludag2005fuzzy,nandakumar2007fingerprint}, we propose different solutions for certain steps to enhance security or accuracy of our algorithm.
In the following, we describe concretely each step and motivate our alternative solutions.

\subsubsection{Vault Encoding}
\label{subsubsec:vault-encoding}

\begin{figure}[t]
	\centering
	\includegraphics[width=\linewidth]{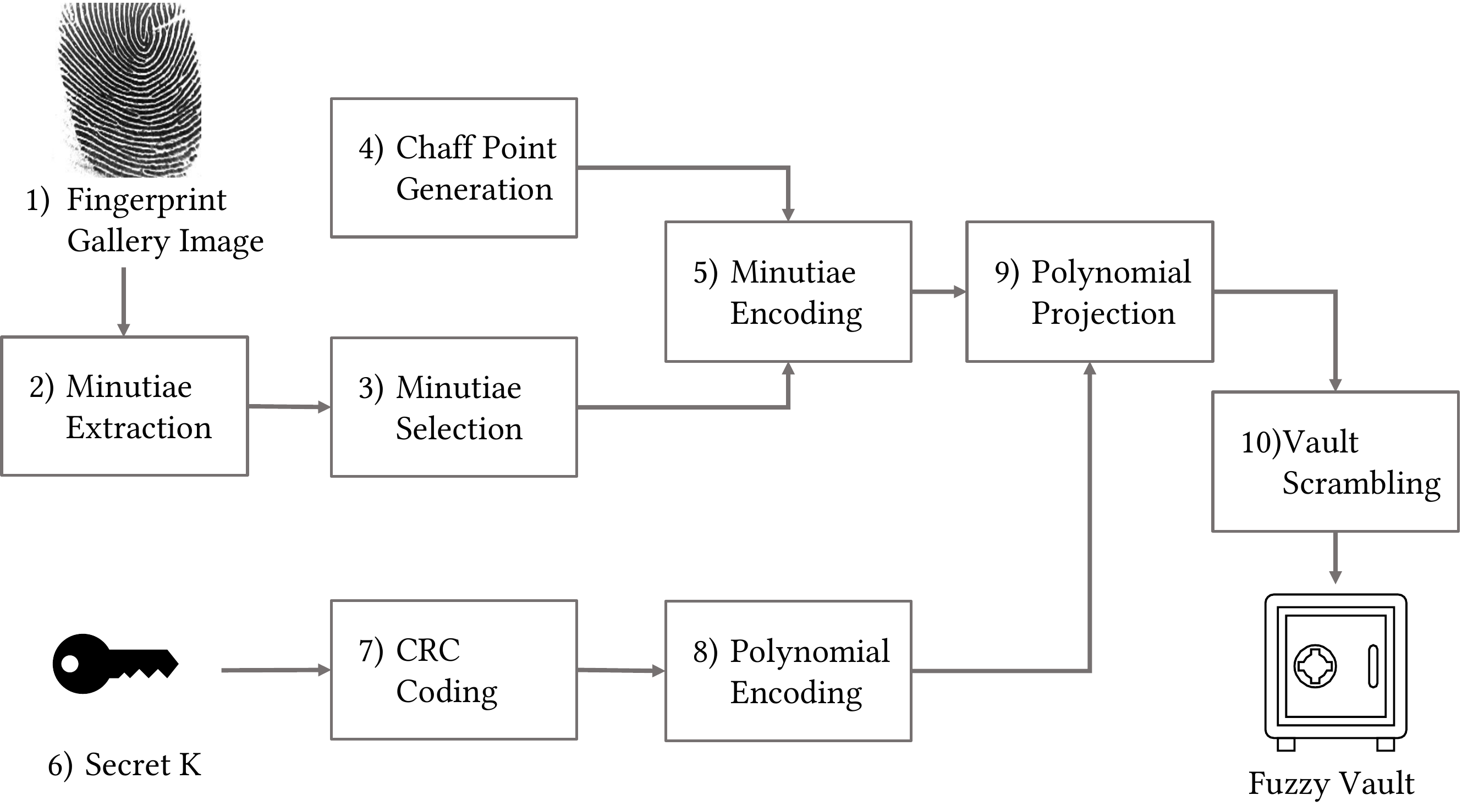}
	\caption{Overview of vault encoding process.}
	\label{fig:vault-encoding}
\end{figure}

The process of vault encoding (steps of \autoref{fig:vault-encoding})
is implemented in our cryptosystem as follows.

\paragraph{1) Fingerprint Gallery Image}
The fingerprint image of the user is given as input to the enrollment algorithm.

\paragraph{2) Minutiae Extraction}
The minutiae are extracted from the given fingerprint image.
These are points that are defined as ridge endings
or ridge bifurcations and they have
three attributes: an $x$- and $y$-coordinate and an
angle $\theta$ which represents the degree of orientation
of a ridge ending or the angle in the middle of a bifurcation.
We call the extracted minutiae from a real fingerprint image genuine minutiae to distinguish them from chaff points or chaff minutiae. 

\paragraph{3) Minutiae Selection}
As the amount of points in the fuzzy vault is assumed to be constant, not all extracted minutiae are used for generating the fuzzy vault but typically only the ones that have good quality in terms of contrast, flow curves and curvature~\cite{ko2007user}.
In our implementation, we rank all extracted minutiae by these quality measures and select the best depending on the chosen parameter for \textit{\#selected minutiae}. Furthermore, we specify a parameter \textit{points distance}, which is the minimum euclidean distance between two minutiae according to the $x$ and $y$ coordinates so that they can be selected for the vault. This reduces the probability of multiple matchings with a single probe minutia.
\paragraph{4) Chaff Points Generation}
In order to obscure the genuine minutiae from a possible attacker, random chaff minutiae or chaff points are generated. In our implementation, we create 10 times as many chaff points as there are genuine minutiae, which yields a reasonable balance between complexity of a brute force attack and performance~\cite{nandakumar2007fingerprint}.
To create a chaff point, we randomly select $x$ and $y$ coordinates within the fingerprint image dimensions and $\theta\in[0,360]$.
We apply the same constraint of a minimum \textit{points distance} threshold as in Minutiae Selection and only accept chaff points 
whose representation is at least half of the smallest genuine minutia selected in the previous step Minutiae Selection to avoid creating points that an attacker
can easily identify as non-genuine.

\paragraph{5) Minutiae Encoding}
For vault encoding we use genuine minutiae extracted from a real fingerprint template and the randomly generated chaff points.
In order to create tuples and polynomial mappings for the fuzzy vault, both minutiae types need to be encoded in the same way. 
In our implementation, we encode the three attributes of each minutia as a bit string of size 32 (11 bits for each $x,y$ coordinate, 10 bits for $\theta$).
We choose this configuration to accommodate typical fingerprint image sizes and allow easy conversion to an unsigned integer for the minutia representation.
The unsigned integers for all genuine minutiae and chaff points are later used for the Polynomial Projection and forming the vault.

\paragraph{6) Secret Generation}
In our algorithm, the secret $K$ is randomly generated as an integer and used for CRC coding. 
The secret size needs to satisfy two constraints. First, its bit length needs to be divisible by 8 to allow smooth conversion to bytes. Second, the length of the secret plus the length of its CRC encoding needs to be divisible by $n + 1$, where $n$ is the polynomial degree, so that the secret can be split equally to polynomial coefficients. 
In our implementation, we determine a suitable secret length $sl$ which satisfies both constraints and generate a random integer between $0$ and $2^{sl} - 1$ for the secret.

\paragraph{7) CRC Coding}
CRC coding of the chosen secret is needed so that an integrity check can be conducted during CRC Error Detection in vault decoding. We choose a 32-bit CRC over a 16-bit CRC
that was used in~\cite{uludag2005fuzzy,nandakumar2007fingerprint} to achieve lower collision probability and hence better reliability. 

\paragraph{8) Polynomial Encoding}
This step is very similar to Minutiae Encoding. The secret and its CRC coding are both converted to a bit string and the CRC coding is then appended to the generated secret. Afterwards, the whole bit string is split into $n + 1$ parts to get the specific coefficient representations of the secret polynomial $p$. The bit string parts are then interpreted as unsigned integers to get the specific coefficients which are needed for Polynomial Projection.

\paragraph{9) Polynomial Projection}
The structure of the fuzzy vault is a set of $(X,Y)$ tuples, where $(X,Y)$ specifies a point in a coordinate system.
The first element of a vault tuple, $X$, contains a genuine minutia or chaff point, which we also call the vault minutia. 
The second element, $Y$, is generated differently for the two minutiae types.
Every genuine minutiae representation $X$ is mapped on the secret polynomial $p$ and the resulting value is saved as the second element of the tuple, i.e., $Y=p(X)$. 
The polynomial mapping is conducted in a Galois field so that exact polynomial interpolation in vault decoding is possible. 
The chaff points are then randomly mapped to a number in the possible result space of the polynomial projection, so that they do not lie on the polynomial.
This mapping yields the second element of the tuple, $Y$, where $Y \neq p(X)$.

\paragraph{10) Vault Scrambling}
The fuzzy vault is finalized by shuffling all vault tuples so that the tuples with genuine minutiae cannot be distinguished from the ones containing chaff points.
\subsubsection{Vault Decoding}
\label{subsubsec:vault-decoding}
\begin{figure}[t]
	\centering
	\includegraphics[width=\linewidth]{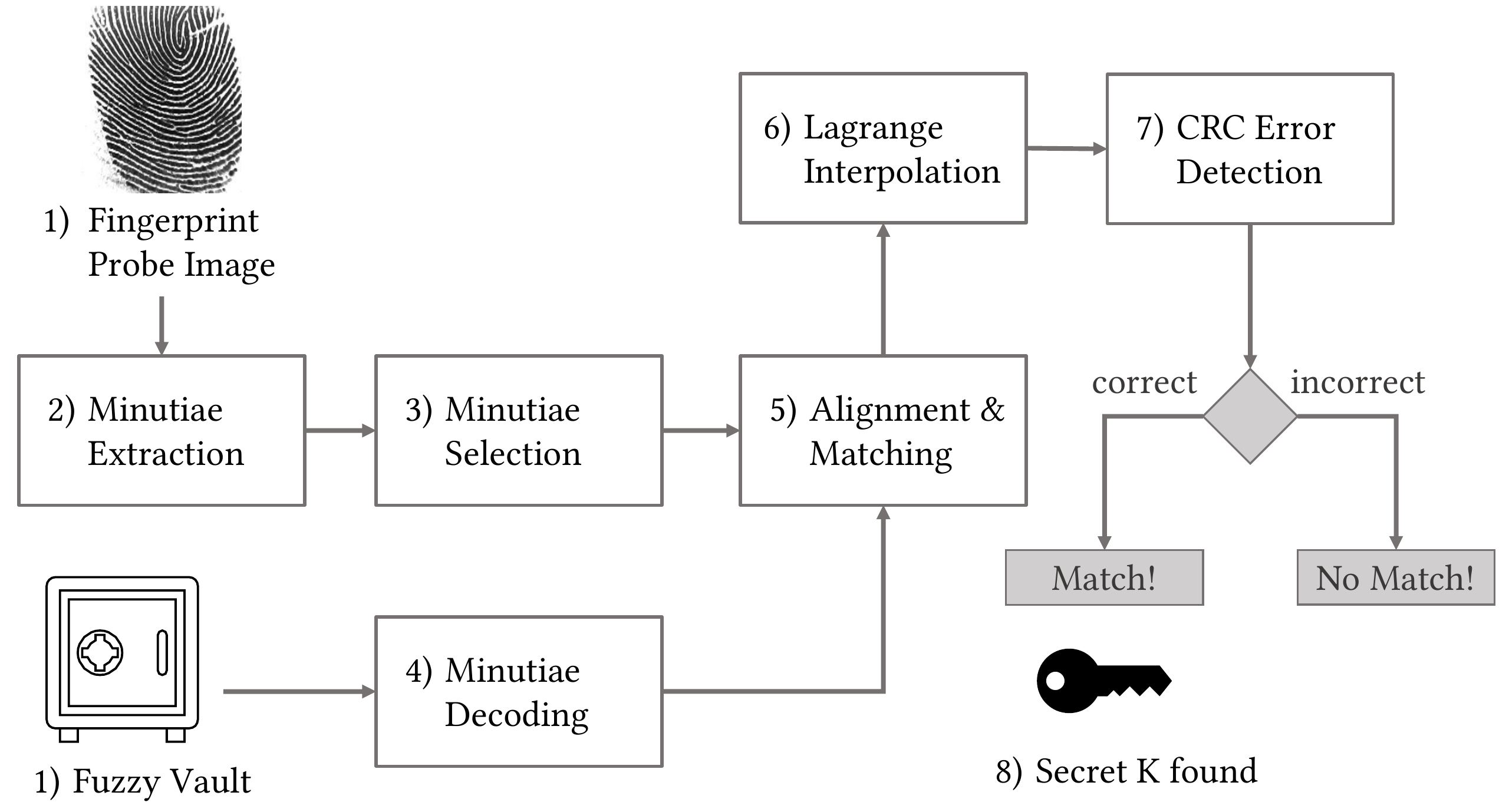}
	\caption{Overview of vault decoding process.}
	\label{fig:vault-decoding}
\end{figure}

The process of vault decoding is illustrated in \autoref{fig:vault-decoding}
and implemented as follows.

\paragraph{1) Fingerprint Probe Image and Fuzzy Vault}
The fingerprint image and the existing fuzzy vault of the (claimed) user is given as input to the verification algorithm.

\paragraph{2) Minutiae Extraction}
The minutiae are extracted from the probe fingerprint image in the same way as in vault encoding.

\paragraph{3) Minutiae Selection}
The best-quality minutiae are selected in the same way as in vault encoding.

\paragraph{4) Vault Minutiae Decoding}
For matching purposes, the vault minutiae, which are the first elements of the vault tuples, need to be decoded to actual minutiae from their representation as unsigned integers. The decoding process is the reverse of step Minutiae Encoding in vault encoding. The unsigned integers are interpreted as bit strings with 32 bits length, from which 
the values of the $x$, $y$, $\theta$ attributes can be retrieved.

\paragraph{5) Alignment and Matching}
When matching two fingerprint captures, variations such as translation and rotation differences are common because having two captures recorded in the exact same way by the fingerprint sensor is extremely unlikely. 
This problem affects the matching of probe minutiae with genuine minutiae in the vault, and to address it,
previous fuzzy vault implementations rely on pre-aligned fingerprint images~\cite{uludag2005fuzzy} or
helper data~\cite{nandakumar2007fingerprint,uludag2006securing}.

In our implementation, we use the approach of geometric hashing~\cite{wolfson1997geometric}.
Geometric hashing is a technique that originated from computer vision to match geometric features against a feature database. 
Chung et al.~\cite{chung2005automatic} applied this method to fuzzy vaults to reduce intra-class variations, 
specifically the discrepancies between different captures of the same finger. 
We choose geometric hashing as our alignment method because it uses only the points in the fuzzy vault. Hence no supplementary public helper data is needed 
like in other global alignment schemes, \eg orientation field-based schemes~\cite{uludag2006securing}, which can leak information about the genuine minutiae. 
Additionally, compared to local alignment approaches, like five nearest neighbor or Voronoi neighbors~\cite{jeffers2007fingerprint}, geometric hashing can achieve higher matching accuracy~\cite{chung2005automatic}.

We implement the basic approach of geometric hashing without an actual hash table. We create a list of lists, which we call geometric table. Both the top list and the sublists have the same length, equal to the number of vault minutiae. Each sublist or element of the geometric table is associated with one specific vault minutia (first element of a vault tuple) as basis and contains all other vault minutiae transformed according to the selected basis. The basis minutia is moved to the origin $(0, 0)$ of a coordinate system, which represents the $x$ and $y$ minutia coordinates on its axes. Then the basis minutia is rotated, so that its orientation points horizontally to the right. The same translation and rotation is then applied to all other vault minutiae. Each vault minutia is chosen as basis once and we align all other vault minutiae to it in its element in the geometric table. We still keep the original vault tuples, which are now referenced by the transformed vault minutiae. We repeat the same procedure for the minutiae extracted from the probe fingerprint and also create a geometric table, where we choose each probe minutia as basis once and transform all others accordingly.
\begin{figure}[t]
	\centering
	\includegraphics[width=\linewidth]{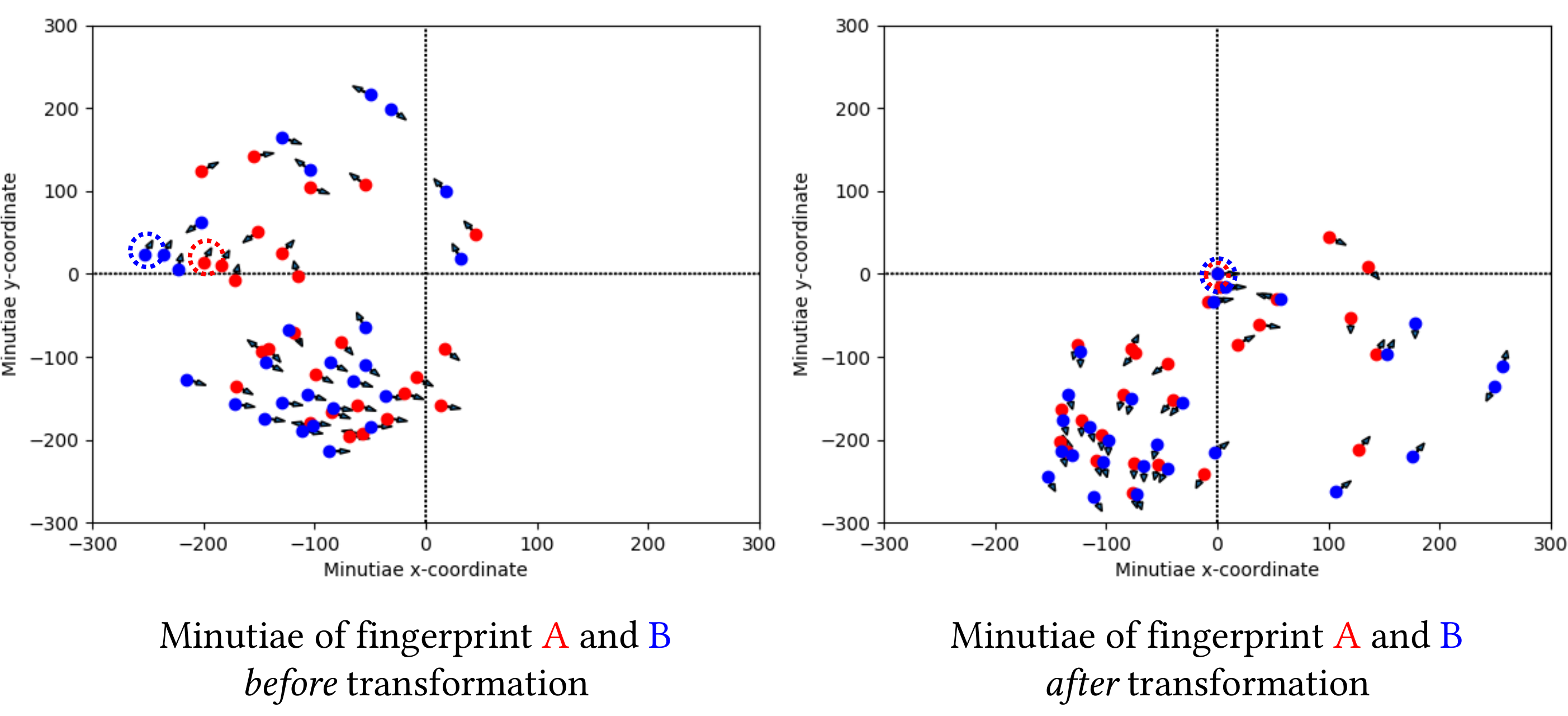}
	\caption{Geometric hashing applied on fingerprint minutiae.}
	\label{fig:geometric-hashing}
\end{figure}

We use these two geometric tables for matching. Our algorithm iterates through all probe minutiae as a basis once and tries to find a similar basis in the geometric table of the fuzzy vault. The purpose of those transformations is to find two minutiae that are selected as basis and are actually the same minutia on a finger. In this case, all other minutiae should be well-aligned and can be matched with small thresholds. In \autoref{fig:geometric-hashing}, we illustrate the transformation with two minutiae that are identical but from two different captures, one from a gallery template found in a vault and the other from a probe template. The two chosen basis minutiae are depicted in red and blue dotted lines, respectively.


After fixing one element of the probe geometric table, we look for a suitable element in the vault geometric table whose basis has similar orientation to the one of the probe geometric table. After finding two suitable bases within the \textit{basis $\theta$ threshold}, every corresponding transformed minutia from the probe is matched against all transformed vault minutiae of the selected basis in the geometric table of the vault. This implies that the minutiae are only compared in a transformed space according to their assigned bases.

To match two minutiae, we use three primary thresholds, \ie for $x$, $y$, $\theta$. If two minutiae satisfy all thresholds, \ie all differences in the attributes are lower than the respective thresholds, the two minutiae are considered to be a match and we place the corresponding vault tuple into a candidate set. This set contains vault tuples which are used for Polynomial Interpolation in the next step.

After all minutiae from two chosen bases have been evaluated, a polynomial interpolation is attempted if the amount of vault tuples in the candidate set is larger than $n + 1$, where $n$ is the polynomial degree. Otherwise, another similar basis is chosen to repeat the minutiae matching. If all probe minutiae have been chosen as basis once and no match has been registered, a failure for unlocking the vault can be reported already at this step.

\paragraph{6) Polynomial Interpolation}
Polynomial interpolation is initiated if the size of the candidate set is larger than $n + 1$ at the end of one run with two given bases. If fewer candidates are found, the secret polynomial cannot be interpolated. 
In case of sufficient vault tuples in the candidate set, all subsets of size $n + 1$ need to be evaluated as not all candidate tuples contain genuine gallery minutiae. For each subset, Lagrange interpolation is conducted in a Galois field $GF(2^{32})$ using $n + 1$ vault tuples, which are interpreted as $(X, Y)$ data points with $X$ and $Y$ being the first and second element of the vault tuple, respectively. 
The resulting interpolated polynomial is passed to the next step. If the CRC error detection reports an error, the algorithm continues to interpolate other subsets 
of the candidate set and eventually proceeds to match vault minutiae with other bases if no subset has given a match.

\paragraph{7) CRC Error Detection}
The interpolated polynomial represents the concatenation of the original secret and its CRC coding of the fuzzy vault. Therefore, we can use CRC as an integrity check and apply the reverse scheme as in step CRC Coding of vault encoding. The coefficients of the polynomial are encoded as a bit string. The last 32 bits, which represent the CRC coding of the actual secret, are cut from the bit string. Afterwards, CRC coding is applied to the secret, \ie the rest of the bit string. If the resulting CRC coding is equal to the 32 bits cut from the representation of the polynomial, then with high probability the correct secret was retrieved and our algorithm reports the successful unlocking of the vault and therefore a match.

\paragraph{8) Correct Secret}
If no error is detected, the correct secret has been retrieved and the user is successfully verified.

\subsection{Implementation Alternatives \& Trade-offs}
\label{subsec:implementation-challenges-contributions}

\subsubsection{Exact Calculations in Galois Field}
\label{subsec:galois-field}
An exact polynomial interpolation is essential in the vault decoding phase for the integrity check to succeed. 
For this, all polynomial projections of the genuine minutiae as well as the polynomial interpolation have to be calculated in a Galois field. 
If the interpolations are conducted without a Galois field, \eg by least-square fitting, the interpolation result will not be exact and thus, the integrity check cannot succeed.

In our implementation, we use the Galois field $GF(2^{32})$ so that all possible minutiae representations, which are 32-bit unsigned integers, can be uniquely represented without quantization. To convert the 32-bit unsigned integer to $GF(2^{32})$, we first transform the unsigned integer to a bit string representation. As a Galois field can be seen as a set of polynomials, we interpret the bit string into coefficients of a polynomial. With 32 bits we get exactly $2^{32}$ elements for $GF(2^{32})$. An irreducible polynomial is found iteratively in our algorithm so that the same irreducible polynomial is calculated for a particular Galois field. 
Operating in $GF(2^{32})$ prevents the need for quantization that existed in previous works~\cite{uludag2005fuzzy,nandakumar2007fingerprint} 
at the cost of increased computational complexity.

\subsubsection{Alignment with Geometric Hashing}
\label{subsec:alignment-geometric-hashing}
The security benefit of using geometric hashing over other alignment solutions has been justified earlier.
Increased security and relative simplicity of implementation may
come at the cost of runtime performance though, as a lot of bases and minutiae need to be matched. 
By introducing parameter \textit{$\theta$ basis threshold} and only considering as bases those minutiae in the vault that have a similar orientation to the minutiae received from the probe template, we can considerably decrease the matching possibilities and therefore the matching runtime. Using such a threshold is a reasonable assumption as typically the fingerprint captures do not differ by more than 10 to 20 degrees. In our experiments (\autoref{sec:results}), we observe that the total runtime without \textit{$\theta$ basis threshold} is more than three times higher compared to using a \textit{$\theta$ basis threshold} of $10$ degrees.

\subsubsection{Minutiae Matching Thresholds}
\label{subsec:matching-thresholds}

The specification of minutiae matching thresholds \textit{points distance}, \textit{x}, \textit{y} and \textit{$\theta$ thresholds} has a significant impact on the accuracy of our algorithm. Optimal parameters depend greatly on the input fingerprint database as well as the particular use case and its requirements for usability and security. We have analyzed the impact of different parameters extensively for the FVC2006 DB 2A~\cite{cappelli2007fingerprint} in \autoref{sec:results}.

A sensitive point in choosing thresholds is the possibility that a probe minutia could match multiple gallery minutiae in the vault. This is possible if two gallery minutiae lie near to each other and the probe minutia is also close, when matched in the transformed space defined by two suitable bases. It would not matter if chaff points are by chance matched as they would not provide information to retrieve the secret. In the worst case, an attacker would need fewer probe minutiae to match multiple correct minutiae.

We could avoid such a scenario by only allowing one probe minutia to match one vault minutia. However, if the probe and gallery fingerprint templates only have few matchable minutiae and a few of the probe minutiae are matched with chaff points, then we would have an incorrect match and lower accuracy. Because the order of matching minutiae is random in our algorithm, we could even get inconsistent results depending on whether a chaff point or a genuine minutia matches with a probe minutia. The addition of parameter \textit{points distance} can prevent such a scenario if the thresholds are chosen accordingly. This would mean that all vault minutiae need to be separated by at least the \textit{points distance}, which needs to be larger than the minutiae matching thresholds.

However, in practice this approach also does not work as the higher \textit{points distance} is, the more difficult it gets to find chaff points, as a newly generated chaff point needs to keep the \textit{points distance} to all previous vault minutiae. If we decrease the minutiae matching thresholds instead, the accuracy of our algorithm drops considerably as the matching becomes too conservative and few matches can be reported. In our implementation, we accept the risk that an attacker gains a small advantage by possibly matching multiple correct minutiae with a single probe minutia. Our algorithm adds all vault tuples to the candidate set whose vault minutia is within the minutiae matching thresholds of at least one probe minutia of the user who wants to authenticate.

\subsubsection{Generation of Candidate Subsets}
In order to extract the polynomial coefficients and therefore the secret, subsets of size $n + 1$ need to be found from the whole candidate minutiae set, which is typically much larger. We implemented three different approaches to generate candidate subsets in our system, each with its own benefits and drawbacks:
\begin{itemize}
	\item Iterative Selection: With iterative selection the subsets are created in order. This means that only one element is swapped out at a time. If the selected first subset contains very few genuine minutiae, a lot of iterations are needed to eventually swap out the chaff minutiae tuples. Our experiments show a much slower average time for subset generation, by almost a factor of 10, compared to the other two approaches. However, this method guarantees a deterministic result as all possible subsets are considered.
	\item Random Generation: In random generation, all possible subsets are created and iterated after shuffling. This typically avoids the situation described in iterative selection and leads to smaller average times. This method also guarantees deterministic results as all possible subsets are traversed. Nevertheless, this approach has a big drawback in having to generate all subsets upfront. If the candidate minutiae set is very large, such as more than 35 elements, and the chosen \textit{polynomial degree} is higher than 12, there could be a memory overflow on a normal computer with 8 GB of RAM. Moreover, the generation of the subsets upfront takes a fixed amount of time, independent on when a match is found.
	\item Random Selection: In random selection, the subsets are randomly chosen on the fly. In each iteration, the algorithm randomly chooses a subset out of the candidate set. The number of iterations is set to be the same as the total number of possible subset combinations in the candidate set. This typically leads to small average execution times and does not incur memory problems. In our experiments, the random selection method is 10 to 25\% faster than the random generation technique. However, this method has the drawback of being non-deterministic, as not all possible subsets need to be selected.
\end{itemize}
In our implementation, we recommend random selection for the generation of candidate subsets. The probability of having ambiguous results is empirically very small and random selection yields the fastest runtimes.

\subsubsection{CRC versus SHA}
\label{subsec:crc-vs-sha}
Most fuzzy vault-based algorithms use CRC as an integrity check~\cite{uludag2005fuzzy,nandakumar2007fingerprint}. We consider SHA as a potentially better alternative as fewer collisions can be expected. For example, SHA-256 maps to 256 bits while CRC-32 only maps to 32 bits, \ie integrity check with SHA-256 is more reliable.

However, there is a trade-off with security when using a hash function with a large digest. In our current algorithm, we operate in a Galois field $GF(2^{32})$ and map each 32-bit unsigned integer to exactly one element in the field. This implies that each of the coefficients in the secret polynomial can only be 32 bits, so that we can convert each of them into one specific $GF(2^{32})$ element. If we, for example, choose SHA-256 as our integrity check, we would already need 8 coefficients just to encode the SHA-256 digest. Therefore, at least a polynomial degree of 8 is needed to have 9 coefficients available. With a polynomial degree of 8, the secret would only be 32 bits. Of course, choosing a very small secret also brings the risk of a potential attacker trying to brute-force the secret directly. If the attacker finds the 32-bit secret, the SHA-256 digest can be calculated and thus the secret polynomial. Therefore, the attacker only needs to check the vault pairs to see which pairs map onto the polynomial to extract the genuine minutiae.

\section{Distributed Access Control}
\label{sec:distributed-fingerprint-application}
Our distributed access control application serves as a proof of concept for the applicability of fingerprint-based authentication with
fuzzy vault in a real-world setting. We present a prototype setup using commercial fingerprint sensors and a cloud environment.

\subsection{Setup and Services}
\begin{figure}[t]
	\centering
	\includegraphics[width=\linewidth]{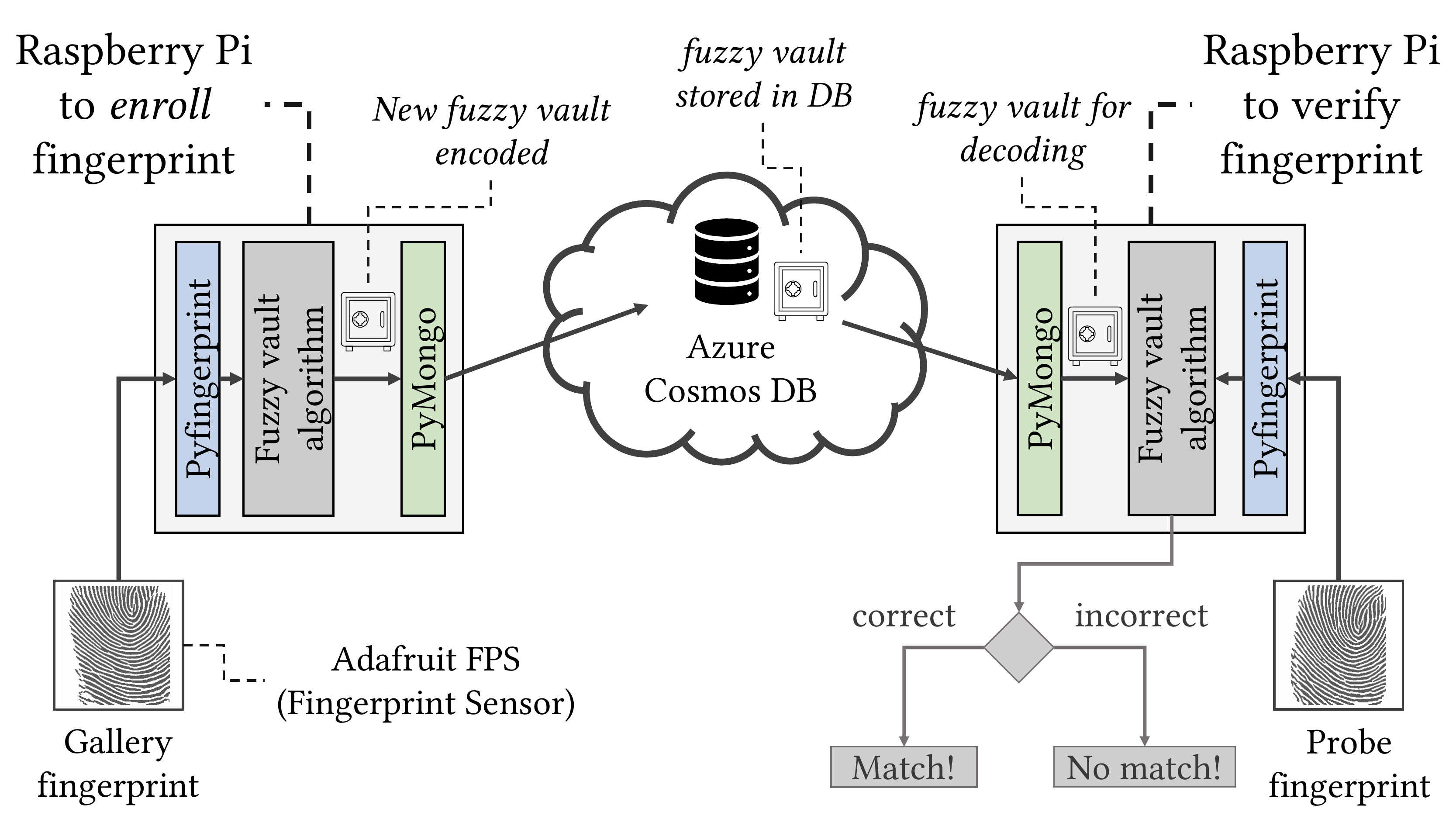}
	\caption{Distributed access control application to enroll and verify fingerprints.}
	\label{fig:distributed-setup}
\end{figure}

We use the following hardware, libraries and services to build our distributed system:
\begin{itemize}
	\item Adafruit Fingerprint Sensor~\cite{website:adafruit-fingerprint}: We use the Adafruit optical sensor to capture fingerprint images from users. 
	We do not use the in-built matching algorithms of the sensor. 
	\item Raspberry Pi~\cite{website:raspberry-pi}: Raspberry Pis serve as our local computers and run our algorithm including vault encoding and decoding 
	(\autoref{subsec:fingerprint-authentication-algorithm}) locally.
	\item Pyfingerprint~\cite{website:pyfingerprint}: We use the Pyfingerprint library to connect Adafruit fingerprint sensors with Raspberry Pis.
	\item Azure Cosmos DB~\cite{website:azure-cosmos-db}: The Azure Cosmos DB serves as our database and storage in the cloud. 
	\item PyMongo~\cite{website:pymongo}: To access Azure Cosmos DB, we use PyMongo and the MongoDB API provided by Cosmos DB.
\end{itemize}
The data flow between the hardware, libraries and services is illustrated through arrows in \autoref{fig:distributed-setup}. In our prototype system,
we use two Raspberry Pis; one is used for enrolling and the other for verifying fingerprints. Both local computers have the same capabilities and can enroll and verify fingerprints interchangeably according to the system model of \autoref{subsec:system-model}.

\subsection{Application Features}
Our application has two features, \ie enrolling a fingerprint to the database on Microsoft Azure and verifying if a fingerprint is registered. For example, a possible use case would be access control for multiple buildings. In this case, the user can enroll his fingerprint once at any building to gain access to all related buildings afterwards. Of course, the enrollment process has to be restricted to authorized people in a real-world setting.

\subsubsection{Enroll Fingerprint}
\label{subsubsec:enroll-fingerprint}
To enroll a fingerprint, the user is prompted for an ID number, which is later used to verify the fingerprint. 
After issuing an ID number, the user can enroll a fingerprint by placing a finger on the Adafruit fingerprint sensor. The image is downloaded to the connected Raspberry Pi and the minutiae are extracted. If the amount of extracted minutiae is smaller than the fixed parameter of \textit{\#selected minutiae}, the user is prompted to rescan the finger. Typically the threshold is set to be between 25 and 50, while a finger usually has 50 to 80 minutiae. Hence, a rescan is normally only needed if the finger is not scanned correctly.

If a suitable fingerprint image with enough minutiae is captured, a fuzzy vault is created according to vault encoding.
The previously defined ID number is attached to the fuzzy vault and the whole structure is serialized to a JSON object using Python dictionaries and sent to Azure Cosmos DB with PyMongo. The JSON object contains an internal object ID, the ID number provided by the user and an array of $(X, Y)$ vault tuples. 
After the database confirms the reception of the vault, the local vault as well as the fingerprint image and template are deleted. This concludes a successful fingerprint enrollment.

\textit{Note}: Besides fuzzy vault, the whole geometric table could be stored in Azure Cosmos DB to avoid computing the geometric table 
on the local computer after receiving the vault for decoding. This approach was infeasible with Azure Cosmos DB due to the document size limit of 2 MB
(the geometric table is in practice larger than 3 MB), but can be considered if no such constraints apply.

\subsubsection{Verify Fingerprint}
\label{subsubsec:verify-fingerprint}
To verify a fingerprint, the user is prompt-ed for an ID number first, so that the corresponding fuzzy vault can be retrieved from the database. 
The fuzzy vault in JSON format is fetched from Cosmos DB and deserialized for vault decoding.
The user is asked to place his finger on the Adafruit fingerprint sensor, so that a probe fingerprint with sufficient minutiae according to the parameter \textit{\#selected minutiae} can be captured.

After a suitable fingerprint image is scanned, vault decoding is run with the fuzzy vault. If the vault is successfully decoded, meaning that the CRC of the secret matches, a match is reported. Otherwise, the system reports an authentication failure. The probe fingerprint template and the corresponding image are then deleted from the local computer.

\textit{Note}: Throughout the paper, we consider user verification rather than identification (N-to-1 matching) because identification with large fingerprint
databases poses critical security risks and should thus be avoided~\cite{prabhakar2003biometric}.
Instead, we use an ID number and fingerprint verification, which only matches one vault with one probe fingerprint. 
The ID number serves as a PIN and does not necessarily need to be unique. 
As long as there are not too many same entries for a specific ID, fingerprint identification can be used for the same ID numbers. 
In this case, a match would be attempted for all fuzzy vaults that are assigned to the same ID and only one needs to match. 
If ID numbers are restricted to be unique, fingerprint verification is guaranteed and better security can be achieved.

\thispagestyle{empty}
\section{Security Analysis}
\label{sec:attack-genuine-vault-pairs}

Security and privacy guarantees strongly depend on where the assets, like fingerprint templates, are used and stored. 
According to our threat model (\autoref{subsec:threat-model}), in our distributed system, fingerprint enrollment and verification
run on local computers within a secure execution environment and the cloud server where the vaults are stored is honest but curious.
Computer storage outside the secure execution environment and cloud storage are not trusted.

By creating the fuzzy vault locally, we do not need to transfer the sensitive fingerprint image to the server. 
The image can be deleted right after vault creation and so can the probe fingerprint that is used for verification. 
However, as the fuzzy vault is stored and transferred in unencrypted form, an attacker can potentially steal a fuzzy vault and attempt to unlock it. 
In this section, we describe an offline brute-force attack which iterates through all possible vault pair subsets and attempts to interpolate the secret polynomial. If a subset is found to correctly interpolate the secret, the attacker can generate a fingerprint template with the first element of the vault pairs in the respective subset to unlock the fuzzy vault in the online algorithm. The attack is analyzed theoretically in \autoref{subsec:genuine-vault-pairs-theoretic} and empirically in \autoref{subsec:genuine-vault-pairs-empirical}.

Note that, alternatively, the attacker can attempt to generate similar minutiae to match the genuine vault pairs and therefore find sufficient genuine minutiae to interpolate. This method is harder for an attacker to achieve and for brevity, its analysis is omitted.
\vspace{-0.3cm}

\subsection{Theoretical Analysis}
\label{subsec:genuine-vault-pairs-theoretic}
For a secret polynomial $p(x)$ with polynomial degree $n$, a subset of vault pairs of size $n + 1$ is needed for its interpolation. In this case all $n + 1$ vault pair elements need to contain genuine gallery minutiae assuming that the integrity check with CRC works correctly and chaff points do not lie on the secret polynomial.

We assume that there are in total $v = g + c$ vault pairs in the fuzzy vault, where $g$ is the number of genuine gallery minutiae and $c$ is the number of chaff points defined at vault creation. This gives the attacker $v_s = \binom{v}{n + 1}$ possible vault pair subsets of size $n + 1$ overall. Out of those sets, $g_s = \binom{g}{n + 1}$ subsets contain genuine gallery minutiae entirely, which can be used to interpolate the secret polynomial correctly. 
This leaves $c_s = v_s - g_s$ chaff pair subsets that can contain genuine vault pairs but include at least one chaff vault pair. If only one element of the candidate subset of size $n + 1$ is a chaff vault pair, the interpolation of the secret polynomial fails due to the chaff vault pair not lying on the secret polynomial.

The attacker has no information about which vault pairs contain genuine gallery minutiae and therefore he/she can only randomly choose one subset after another to interpolate. For $1 \leq i \leq c_s$, let $C_i$ be a random variable that returns 1 if the $i$th chaff subset is selected before selecting any of the genuine subsets, otherwise it returns 0.
The expected number of attempts an attacker needs to find a suitable subset for correct interpolation of the secret can be calculated as follows: 
\begin{align}
\mathbb{E} (g_s, c_s) = & \quad 1 + \sum_{i=1}^{c_s}\mathbb{E}(C_i) = 1 + c_s \cdot \frac{1}{g_s + 1} \nonumber \\
=& \quad\frac{g_s + c_s + 1}{g_s + 1} = \frac{v_s + 1}{g_s + 1}\label{eq:expectation-subsets}
\end{align}
The first term ($1$) represents the terminating first genuine subset that is selected. $\frac{1}{g_s + 1}$ is the probability that the $i$th chaff subset is selected before any of the genuine subsets, where ${g_s + 1}$ represents the total number of genuine subsets and the $i$th chaff subset. 
We repeat this calculation for each chaff pair subset.

To calculate the expected time of a possible attacker to unlock a fuzzy vault by subset interpolation randomly, we multiply the expected time to choose a genuine subset with the average Lagrange interpolation time per subset $l(n)$: 
\begin{equation}
{\mathbb{E}(g_s, c_s) \cdot l(n) = \frac{v_s + 1}{g_s + 1}} \cdot l(n) \label{eq:expectation-time-subsets}
\end{equation}
Note that the average interpolation time increases linearly with the polynomial degree $n$. 

In the following, we summarize the impact of parameters $g$, $c$ and $n$ on the expected time an attacker needs to unlock the secret. To better show the security impact, we represent the expected value $\mathbb{E}(g_s, c_s)$ with expanded binomial coefficients:
\begin{equation}
{\mathbb{E}(g_s, c_s) = \frac{v_s + 1}{g_s + 1} = \frac{\binom{v}{n + 1} + 1}{\binom{g}{n + 1} + 1} = \frac{\frac{v\,!}{(n+1)\,!(v-n-1)\,!} + 1}{\frac{g\,!}{(n+1)\,!(g-n-1)\,!} + 1}} \label{eq:expectation-subsets-binomial}
\end{equation}

\begin{itemize}
	\item By increasing the number of genuine minutiae $g$, the expected value $\mathbb{E}(g_s, c_s)$ decreases as only the denominator grows, hence security decreases. 
	\item By increasing the number of chaff points $c$, $\mathbb{E}(g_s, c_s)$ increases as well, as $v = g + c$ in the nominator increases. Therefore, the attacker needs more attempts to unlock the vault on average. 
	\item By changing the polynomial degree $n$, both binomial coefficients in nominator and denominator change as well. The relations are shown in Eq.~\ref{eq:binom-relations} for $v$, but apply similarly to $g$. In our algorithm, $v$ is an order of magnitude larger than $n$, which implies that when $n$ is increased, $\binom{v}{n + 1}$ increases. $g$ is not as large as $v$ but at least twice as large as $n+1$ in our configurations, so the same relation applies.
	\begin{equation}
	\text{n increases}\left\{
	\begin{array}{@{}ll@{}}
	\binom{v}{n + 1}\ \text{decreases}, & \text{if}\ n+1 \geq \lceil\frac{v}{2}\rceil \\
	\binom{v}{n + 1}\ \text{increases}, & \text{if}\ n+1 < \lfloor\frac{v}{2}\rfloor \\
	\binom{v}{n + 1}\ \text{stays the same}, & \text{otherwise}
	\end{array}\right. \label{eq:binom-relations}
	\end{equation}
	
	Both binomial coefficients in Eq.~\ref{eq:expectation-subsets-binomial} therefore increase with increasing $n$. However, the term $(v-n-1)\,!$ is shrinking faster than the term $(g-n-1)\,!$ as $v = g + c>g$ with $c > 0$. This implies that the fraction with $n$ in the nominator is growing faster than the fraction with $g$ in the denominator, \ie the expected value increases overall.
	As $\mathbb{E}(g_s, c_s)$ increases with increasing $n$, the attacker needs more attempts to find a suitable subset to interpolate the correct secret, thus enhancing security.
\end{itemize}

\subsection{Empirical Analysis}
\label{subsec:genuine-vault-pairs-empirical}
We estimate how many attempts and how much time an attacker needs to unlock a fuzzy vault with the strategy mentioned above by using typical values for the parameters $g$, $c$ and $n$ and a realistic average time $l(n)$ for a single interpolation. $l(n)$ is measured for our implementation of \autoref{sec:distributed-fingerprint-application} using a normal server instance. We present two parameter configurations in \autoref{tab:practical-estimation}. Although both configurations look almost identical, the difference in bit security is significant.
\begin{table}[t]
	\centering
	\resizebox{0.75\columnwidth}{!}{	
	\begin{tabular}{|l|l|l|}
		\hline
		\textbf{Parameter} & \textbf{Configuration 1} & \textbf{Configuration 2} \\ \hline
		$g$ & $35$ & $35$ \\ \hline
		$c$ & $300$ & $300$ \\ \hline
		$v$ & $335$ & $335$ \\ \hline
		$n$ & $8$ & $12$ \\ \hline
		$l(n)$ & $0.01$ s & $0.01$ s \\ \hline
	\end{tabular}}
	\caption{Parameters of empirical security analysis.}
	\label{tab:practical-estimation}
\end{table}

For configuration 1, Eq.~\ref{eq:expectation-subsets} yields approximately $1.86 \cdot 10^{9}$ expected attempts for an attacker to correctly interpolate the secret. This roughly corresponds to a 30-bit security level as $1.86 \cdot 10^{9} \approx 2^{30}$.
By increasing the polynomial degree to 12 as shown in \autoref{tab:practical-estimation}, we get a completely different result. We keep the average subset interpolation time unchanged, as in practice the difference in time between polynomial degree $8$ and $12$ is negligible. For configuration 2, the expected number of attempts increases to approximately $6 \cdot 10^{13}$, which roughly corresponds to a 46-bit security level.

Note that a 46-bit security level is generally not considered secure nowadays. One possibility to enhance the security level is to use multiple fingers to create multiple fuzzy vaults that all need to be matched in order to authenticate. Another possibility is to increase the number of chaff points or the polynomial degree, which will have an impact on the runtime performance and accuracy. The right balance between security and usability needs to be found depending on each particular use case.
\section{Evaluation} \label{sec:results}
This section presents an evaluation of our biometric cryptosystem implementation. 
\autoref{subsec:experimental-setting} describes the experimental setup. In
\autoref{subsec:eval-fuzzy-vault-algorithm} we evaluate the fuzzy vault algorithm using a public fingerprint database and compare it to widely-used algorithms for fingerprint recognition. 
In \autoref{subsec:evaluation-param-impact} we analyze the impact of parameter choices on security and usability, and in \autoref{subsec:eval-distributed-app}
we present results specific to the distributed access control application.

\subsection{Experimental Setup}
\label{subsec:experimental-setting}
\subsubsection{Data Sets and Protocols}
We run the authentication algorithm of \autoref{subsec:fingerprint-authentication-algorithm}
against the Fingerprint Verification Competition 2006 (FVC2006) database 2A~\cite{cappelli2007fingerprint} with images from optical sensors. 
These images are similar to images obtained by the Adafruit optical fingerprint sensor and have the highest image resolution across all FVC2006 databases. 
Database 2A consists of fingerprint images from 140 fingers with 12 captures each, resulting in 1'680 pictures in total.
For the experiments of \autoref{subsec:eval-distributed-app} we do not use a fingerprint database, as the system is tested with people enrolling and verifying their fingerprints on an Adafruit FPS.
The specifications of the FVC2006 database 2A and the Adafruit FPS are listed in \autoref{tab:fvc2006-adafruit-specs}.

\begin{table}[t]
	\centering
		\resizebox{0.85\columnwidth}{!}{	
	\begin{tabular}{|l|l|l|}
		\hline
		& \textbf{FVC2006 DB 2A}      & \textbf{Adafruit FPS} \\ \hline
		\# fingers           & 140                         & 10                    \\ \hline
		\# captures per finger & 12                          & 5                    \\ \hline
		Sensor                      & BiometriKa (optical) & Adafruit FPS          \\ \hline
		Image size                  & 400 x 560 px                  & 256 x 288 px           \\ \hline
		Image resolution            & 569 dpi                     & 96 dpi          \\ \hline
	\end{tabular}}
	\caption{FVC2006 DB 2A and Adafruit FPS specifications.}
	\label{tab:fvc2006-adafruit-specs}
\end{table}

For our experiments with FVC2006, we consider three different protocols. The \textit{all vs all protocol} analyzes every possible match in database 2A. The \textit{FVC protocol} and \textit{1vs1 protocol} are often used in literature to evaluate FVC databases. For example, Ferrara et al.~\cite{ferrara2012noninvertible} evaluated their P-MCC approach using both protocols against FVC2006 2A. 
Since the results of all protocols show very similar trends, we present only the results for the \textit{FVC protocol} for brevity.

In the \textit{FVC protocol}, to evaluate the FNMR, each fingerprint capture is compared against the remaining captures of the same finger. The FMR is determined by comparing the first template of each fingerprint against all other first templates of the remaining fingers. Symmetric comparisons are not executed. If $T1$ is compared against $T2$, $T2$ is not matched against $T1$. This results in 9'240 total matches for the evaluation of the FNMR and 9'730 matches for the FMR, respectively, in FVC2006 2A.

\subsubsection{Software \& Hardware Configuration}
\label{subsec:experiment-setup}
To evaluate the accuracy (\ie FMR and FNMR) and runtime performance of our fuzzy vault algorithm, we use server instances with Intel i5-3470 quad-core processors clocked at 3.4 GHz and run against the FVC2006 database 2A. Our algorithm runs in a Docker container~\cite{website:docker} using PyPy~\cite{website:pypy}, which is a just-in-time compiler for Python. For the evaluation of the algorithm, the fingerprint images are preprocessed with MINDTCT from NBIS~\cite{ko2007user} to perform minutiae extraction and quality assessment and to provide the minutiae templates as input to the algorithm.
Additional library dependencies include the Python binascii library~\cite{website:python-binascii} for CRC coding and Galoistools from Sympy~\cite{website:sympy-galoistools} 
for calculating polynomial mappings and performing Lagrange interpolation in a Galois field.
For future evaluation, the source code of our implementation can be downloaded from~\cite{oss}.

For the distributed application, our algorithm runs on a Raspberry Pi 3 with an ARM Cortex A53 CPU running at 1.2 GHz. Two Raspberry Pis are connected to equal Adafruit fingerprint sensors and the Azure Cosmos DB. MINDTCT~\cite{ko2007user}, binascii~\cite{website:python-binascii} and Galoistools~\cite{website:sympy-galoistools} are integrated in the application as before.


\subsection{Evaluation of Authentication Algorithm}
\label{subsec:eval-fuzzy-vault-algorithm}

\begin{figure}[t]
	\centering
	\includegraphics[width=\linewidth]{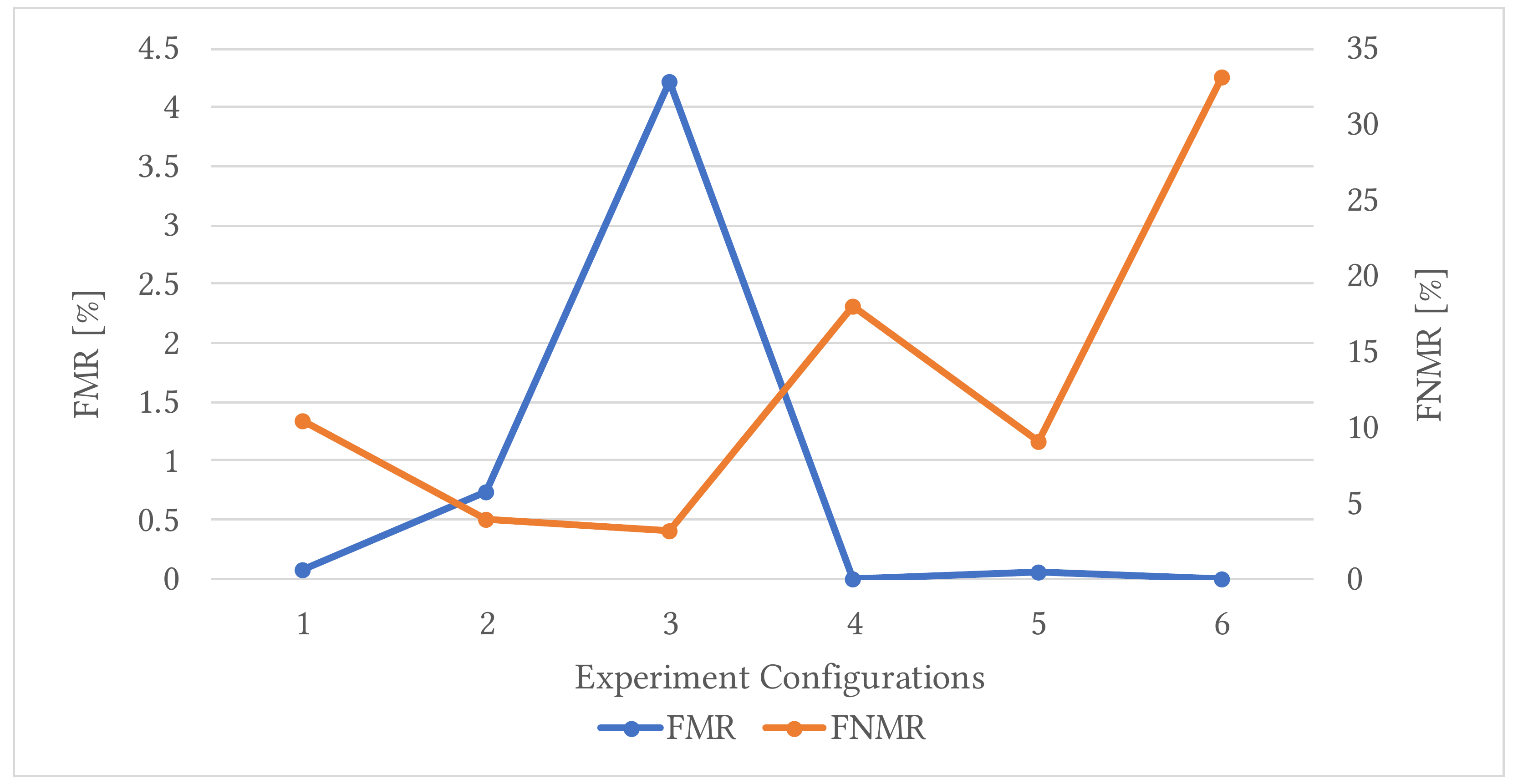}
	\caption{FVC protocol: Accuracy.}
	\label{fig:exp-fvc-accuracy}
\end{figure}

For our experiments with the \textit{FVC protocol}, we consider various parameter configurations as listed in \autoref{tab:exp-fvc-configuration}. 
The considered parameters are the \textit{polynomial degree} $n$, the \textit{\#genuine minutiae} $g$ and \textit{\#chaff points} $c$, \textit{points distance} $pd$,
the minutiae matching thresholds $x_{thres}$, $y_{thres}$,	$\theta_{thres}$ and the \textit{$\theta$ basis threshold} $\theta\text{-basis}_{thres}$.
We only list configurations of interest. Certain thresholds are the same in the shown configurations and have been selected due to their good performance. 
\begin{table}[t]										
	\centering
	\resizebox{.9\columnwidth}{!}{	
		\begin{tabular}{|l|l|l|l|l|l|l|l|l|l|}										
			\hline									
			\textbf{Conf.} &	\textbf{\textit{n}} &	\textbf{\textit{g}} &	\textbf{\textit{c}} &	\textit{pd} &	$x_{thres}$ &	$y_{thres}$ &	$\theta_{thres}$ &	$\theta\text{-basis}_{thres}$	\\ \hline \hline
			1 &	8 &	30 &	340 &	10 &	12 &	12 &	12 &	15	\\ \hline
			2 &	8 &	34 &	300 &	10 &	12 &	12 &	12 &	10	\\ \hline
			3 &	8 &	40 &	300 &	10 &	15 &	15 &	15 &	10	\\ \hline
			4 &	10 &	30 &	300 &	10 &	12 &	12 &	12 &	10	\\ \hline
			5 &	12 &	40 &	300 &	10 &	15 &	15 &	15 &	10	\\ \hline
			6 &	14 &	30 &	300 &	10 &	15 &	15 &	15 &	10	\\ \hline
		\end{tabular}
	}				
	\caption{Configuration of FVC protocol experiments.}									
	\label{tab:exp-fvc-configuration}									
\end{table}
The results of the experiments \wrt accuracy and runtime performance of the authentication algorithm can be found in \autoref{tab:exp-fvc-accuracy-runtime}.
The achieved accuracy of the various configurations is illustrated in \autoref{fig:exp-fvc-accuracy}. 

As can be seen in \autoref{fig:exp-fvc-accuracy}, our algorithm can be tuned to achieve very different FMR/FNMR ratios. 
It always depends on the use case to decide on the desired trade-off between security and usability. For instance,
configuration 3 has the lowest FNMR in all of our experiments with roughly 3\%, however, it would be unacceptable security-wise in most use cases with a 4.22\% FMR. 
Configuration 2 provides an alternative with a relatively low FNMR of 3.89\% but with a much lower FMR of 0.74\%, which is still considered high and therefore only usable in lower-security use cases. Configuration 4 has the smallest FNMR with no single false positive (FMR=0), which makes it suitable for a high-security application, where users are willing to accept some compromise in usability. Configuration 6 shows that by being more conservative with the parameters, here by choosing a higher degree polynomial, the FNMR increases drastically. 
We presume that configurations 1 and 5 achieve a reasonable balance for most practical applications. With these two configurations we achieve an FNMR that is close to the one we get with NBIS BOZORTH3, the fingerprint matching algorithm of NBIS~\cite{ko2007user}, which is a common reference in literature. The respective 
authentication runtime is also acceptable (up to 2.3 sec).

Furthermore, we compare our algorithm to $\text{P-MCC}_{64}$ by Ferrara et al.~\cite{ferrara2012noninvertible}. Their implementation achieves less than 1\% FNMR with 0\% FMR
against FVC2006 2A. For FMR $\leq 0.1\%$, $\text{P-MCC}_{64}$ can achieve an FNMR of less than $0.5\%$. Note, however, that our algorithm provides higher security guarantees as the chance that a randomly selected minutia in a reversed template is real is almost 25\% with $\text{P-MCC}_{64}$~\cite{ferrara2012noninvertible} (compare our reference results in \autoref{subsec:genuine-vault-pairs-empirical}).

We also compare our algorithm to a fuzzy extractor prototype based on the findings by Sutcu et al.~\cite{sutcu2008feature}. Their method revolves around feature vector extraction to obtain minutiae information that is difficult to invert using cuboids. We implemented and ran a basic implementation of the fuzzy extractor against the FVC2006 2A database using the \textit{FVC Protocol} and the same server instances as in \autoref{subsec:experiment-setup}, but using Python3 instead of PyPy3, since this led to better runtime performance. Our results show an FMR of 7.16\%, FNMR of 80.63\% and average total runtime of roughly 7 seconds. Compared to this simple prototype, our algorithm achieves a significantly higher accuracy with a lower runtime.

A direct comparison to previous fuzzy vault implementations~\cite{nandakumar2007fingerprint,uludag2005fuzzy,theodorakis2018secure,li2010alignment} 
was not feasible as their algorithms are evaluated against non publicly available fingerprint databases or older FVC databases. 
Our results on FVC2006 indicate equal or higher accuracy compared to their reported results.
Recall that security is also expected to be better as our approach does not depend on helper data for fingerprint alignment.
\begin{table}[t]								
	\centering
	\resizebox{\columnwidth}{!}{			
		\begin{tabular}{|l|l|l|l|l|l|l|l|}								
			\hline							
			\textbf{Exp} &	\textbf{\textit{n}} &	\textbf{\textit{g}} &	\textbf{FMR [\%]} &	\textbf{FNMR [\%]} &	\textbf{encode [s]} &	\textbf{decode [s]} &	\textbf{total [s]}	\\ \hline \hline
			1 &	8 &	30 &	0.08 &	10.38 &	0.11 &	0.9 &	1.01	\\ \hline
			2 &	8 &	34 &	0.74 &	3.89 &	0.11 &	1.19 &	1.3	\\ \hline
			3 &	8 &	40 &	4.22 &	3.08 &	0.1 &	1.39 &	1.5	\\ \hline
			4 &	10 &	30 &	0 &	18.03 &	0.11 &	0.75 &	0.86	\\ \hline
			5 &	12 &	40 &	0.06 &	9.13 &	0.13 &	2.23 &	2.36	\\ \hline
			6 &	14 &	30 &	0 &	33.16 &	0.15 &	1.84 &	1.99	\\ \hline
		\end{tabular}
	}
	\caption{FVC protocol: Accuracy and runtime performance.}							
	\label{tab:exp-fvc-accuracy-runtime}							
\end{table}
\vspace{0.1cm}
\subsection{Evaluation of Parameter Impact}
\label{subsec:evaluation-param-impact}
\autoref{tab:exp-fvc-accuracy-runtime} indicates that the accuracy in terms of FMR and FNMR and the verification runtime vary significantly for different parameter configurations. 
Here, we evaluate the impact of all parameters on accuracy and runtime (for impact on security, see \autoref{subsec:genuine-vault-pairs-theoretic}). 
We run experiments by fixing all parameters but the one whose impact we analyze. Table~\ref{tab:eval-param-impact} summarizes our experimental results. The impact of each individual parameter is analyzed below to help system developers select appropriate parameters for their applications.\\ \skpline

\noindent
\textbf{Polynomial degree $n$:}
A higher \textit{polynomial degree} increases the minimum amount of minutiae in the candidate set. 
This implies that more genuine minutiae need to be matched and makes matching more conservative, with lower FMR but higher FNMR.
In our experiments, a higher $n$ only marginally increases the total runtime:
although interpolation time increases, there are less interpolations because the threshold for a suitable candidate set also increases.\\ \skpline

\noindent
\textbf{\#genuine minutiae $g$:}
A higher number of genuine minutiae means more selected minutiae in the vault encoding phase and more genuine minutiae found in the vault. 
Therefore, the probability of a match between two fingerprint captures increases, which leads to a higher FMR and lower FNMR.
For higher $g$, the total runtime also increases due to more candidates being found and thus needing more interpolation attempts.\\ \skpline

\noindent
\textbf{\#chaff points $c$:}
The number of chaff points does not impact the accuracy of the algorithm. 
However, the runtime increases with increasing $c$ because more traversals have to be conducted in vault decoding with a larger geometric table.\\ \skpline

\noindent
\textbf{Points distance $pd$:}
If minimum \textit{points distance} is selected to be reasonably small, it affects neither the runtime performance nor the accuracy. Reasonably small in this context refers to similar values as \textit{minutiae matching thresholds}. However, if $pd$ is set too high, the runtime performance drastically decreases as a lot of time is required to find random chaff points that fit the requirement to be at least $pd$ apart from all other vault minutiae. Accuracy is affected considerably only if the $pd$ is so high that not enough good quality genuine gallery minutiae can be selected.\\ \skpline

\noindent
\textbf{Minutiae matching thresholds $x_{thres}$, $y_{thres}$, $\theta_{thres}$:}
The lower these thresholds are, the more conservative the algorithm becomes, with lower FMR but higher FNMR. 
When increasing the thresholds, the opposite effect can be observed. 
A threshold increase also increases the runtime because more minutiae are matched, through which it takes longer to iterate.\\ \skpline

\noindent
\textbf{$\theta$ basis threshold $\theta\text{-basis}_{thres}$:}
We evaluated the impact of changing \textit{$\theta$ basis threshold} between 10 and 20 degrees.
Since no fingerprint captures in the database actually differ by more than 15 degrees,
increasing \textit{$\theta$ basis threshold} led to a decrease in FNMR and slight increase in FMR 
when changing $\theta\text{-basis}_{thres}$ within [0,15]. 
However, the runtime increased slightly with increasing $\theta\text{-basis}_{thres}$ because the decoding takes longer due to more possibilities to match a basis minutia and therefore more iterations. If we remove the \textit{$\theta$ basis threshold} by setting it to 360 degrees so that all bases match, the runtime increases dramatically. In our experiments, we observe a total runtime that is more than three times higher than with a $\theta\text{-basis}_{thres}$ of 10 degrees.
\begin{table}[t]										
	\centering
	\resizebox{\columnwidth}{!}{	
		\begin{tabular}{|l|l|l|l|}									
			\hline
			\textbf{Increase of Parameter} & \textbf{FMR} & \textbf{FNMR} & \textbf{Runtime} \\ \hline									
			\textit{polynomial degree n} & lower &	higher & slightly higher \\ \hline
			\textit{\#genuine minutiae g} & higher & lower & higher	\\ \hline
			\textit{\#chaffpoints c} & unchanged & unchanged & higher	\\ \hline
			\textit{points distance pd} & refer to paragraph &	refer to paragraph & higher \\ \hline
			\textit{minutiae matching thresholds} & higher & lower & higher	\\ \hline
			$\theta\text{-basis}_{thres}$ & slightly higher & lower & slightly higher	\\ \hline
		\end{tabular}
	}				
	\caption{Impact of parameter increase on accuracy/runtime.}									
	\label{tab:eval-param-impact}									
\end{table}
\vspace{0.1cm}
\subsection{Evaluation of Distributed Application}
\label{subsec:eval-distributed-app}
In this section, we focus on the evaluation of the runtime performance of our distributed authentication application.
We do not consider accuracy because the evaluations with the FVC2006 database 2A are more extensive, and representative conclusions have been drawn in previous sections. 
Moreover, the Adafruit FPS provides fingerprint images with lower resolution and quality than the ones in database 2A, thus negatively affecting accuracy. 

To test our algorithm in the distributed setting, we conduct experiments for both enrolling and verifying a fingerprint. We enroll 10 fingers with 5 captures each, which results in 50 invocations of fuzzy vault encoding. Each of the invocations generates an individual fuzzy vault. If a particular fingerprint image has too few minutiae, the application prompts the user for a rescan of the fingerprint. Those rescans are not recorded in the experiments.

\begin{figure}[t]
	\centering
	\includegraphics[width=\columnwidth]{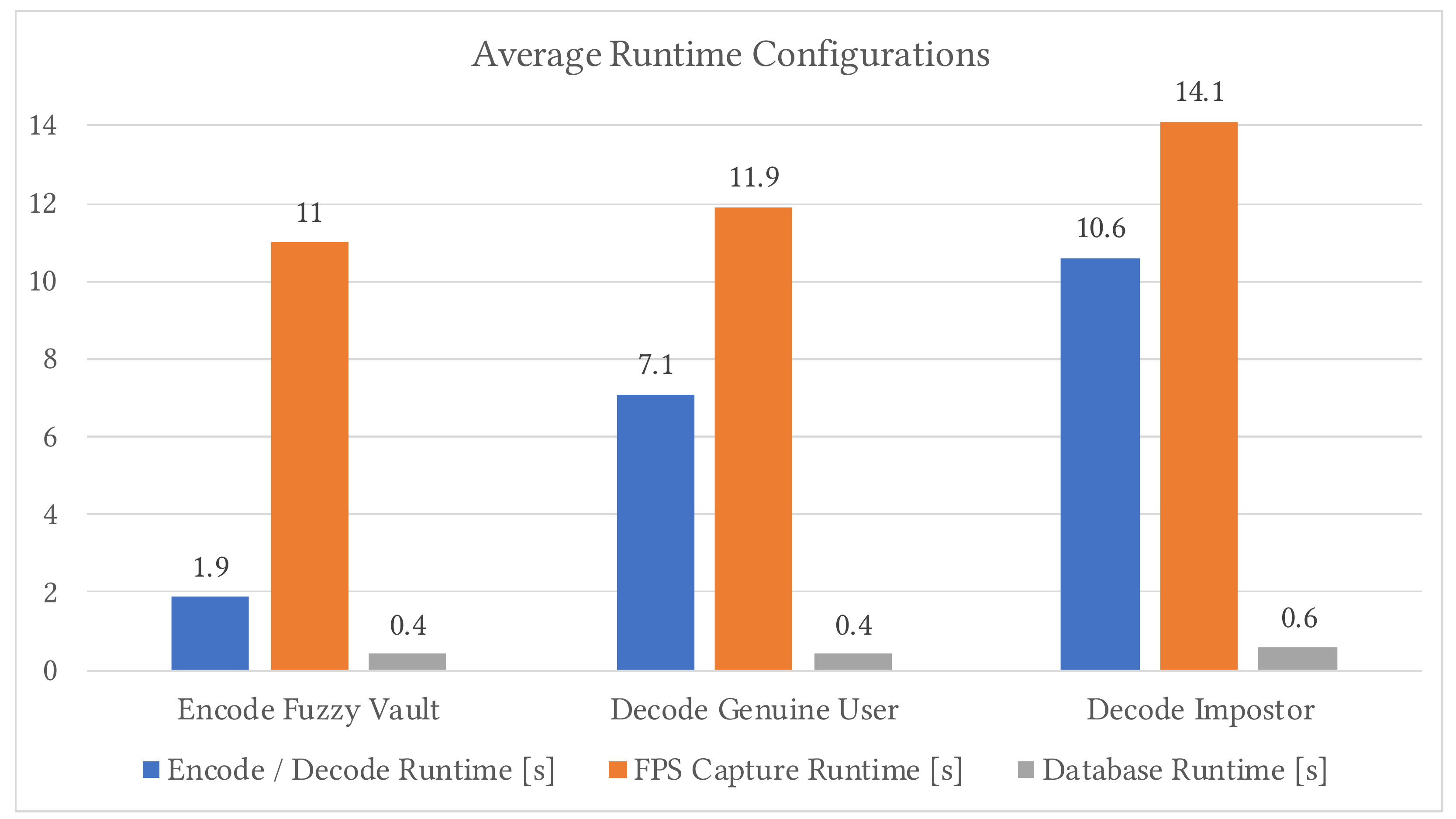}
	\caption{Distributed application runtimes.}
	\label{fig:app-runtime-results}
\end{figure}

To verify the fingerprints, we conduct two different experiments with two configurations. In the first one we verify against each of the 50 fuzzy vaults once, with a capture belonging to the same finger with which the fuzzy vault was generated. This covers the use case of a genuine user trying to authenticate. In the second configuration we also verify against each of the fuzzy vaults once, but use different fingers than were used to create the fuzzy vault. This corresponds to an impostor trying to authenticate. We also evaluate this scenario to show the runtime difference in case of no match. 
The results from our experiments on the distributed application are depicted in Figure~\ref{fig:app-runtime-results}.

Our experiments show that the FPS capture runtime, \ie the time the application needs to download the fingerprint image and run MINDTCT from NBIS~\cite{ko2007user}, represents the largest part of the overall runtime. Given that MINDTCT is very fast on average, we conclude that the FPS capture runtime is constrained by the hardware of the Adafruit FPS which does not allow faster transmission to the Raspberry Pi.
Furthermore, the encoding and decoding runtimes are much higher compared to our results when running the authentication algorithm on a server in previous sections. This is expected due to the limited resources of Raspberry Pi. The difference in decoding runtimes between genuine user and impostor configurations can be justified as generally more iterations are needed in the impostor configuration. When the genuine user configuration is run, the algorithm only runs until a match is found for an average of 14 sec. 
\thispagestyle{empty}
\section{Conclusion} \label{sec:conclusion}

In this work, we designed and built a proof-of-concept biometric cryptosystem based on fuzzy vault that can be used for distributed authentication 
with strong security and privacy guarantees. We presented real-world implementation challenges and proposed solutions and their respective trade-offs. 
The experimental results and the prototype implementation with commercial hardware show that an application of this cryptosystem is feasible in practice,
as it is more secure than widely-used fingerprint recognition algorithms without compromising accuracy or runtime performance. 
At the same time, it improves upon existing cryptosystems in terms of security due to the independence from helper data for fingerprint alignment.
Hence, we consider it a realistic solution for secure and usable distributed authentication.

In future work, we intend to evaluate our access control application on various hardware platforms and use cases and customize it to their requirements 
in order to assess its performance (incl.~communication costs) in industrial settings, based on extensive user studies.


\thispagestyle{empty}

%
\begin{acks}
 We would like to thank Aritra Dhar, Daniele Lain and Srdjan Capkun for their support in the development of the cryptosystem
 as well as Anh Pham and Der-Yeuan Yu for valuable comments on the manuscript.
\end{acks}
\bibliographystyle{ACM-Reference-Format}
\bibliography{literature}

\thispagestyle{empty}

\appendix

\end{document}